\documentclass[aps,prd,preprint,preprintnumbers,unsortedaddress,superscriptaddress,showpacs,nofootinbib]{revtex4-1}

\pdfoutput=1

\usepackage{graphicx,color}
\usepackage{relsize}
\usepackage{slashed}
\usepackage{color}
\usepackage{ulem}
\usepackage{tabu}

\usepackage{amsmath}
\usepackage{amssymb}
\usepackage{amsthm}
\usepackage{mathrsfs}
\usepackage{graphicx}
\usepackage{epstopdf}
\usepackage{fancyhdr}
\usepackage{array}
\usepackage[all]{xy}
\usepackage{eufrak}
\usepackage{euscript}
\usepackage{enumerate}
\usepackage{slashed}
\usepackage{hyperref}
\usepackage{subfigure} 
\usepackage{epstopdf} 
\usepackage{mathtools}

\hypersetup{pdftex,colorlinks=true,linkcolor=blue,citecolor=blue,menucolor=black,urlcolor=blue,filecolor=blue}

\newcommand{\bK}{\bar{K}}

\begin{document}

\title{Unveiling the $K_1(1270)$ double-pole structure in the 
$\bar{B}\to J/\psi \rho\bK$ and $\bar{B}\to J/\psi \bK^*\pi$ 
decays}

\author{J.~M.~Dias}
\email{jorgivan.mdias@gmail.com}
\noaffiliation

\author{G.~Toledo}
\email{toledo@fisica.unam.mx}
\affiliation{Instituto de F\'isica, Universidad Nacional Aut\'onoma de M\'exico, 
AP 20-364, Ciudad de M\'exico 01000, Mexico.}

\author{L.~Roca}
\email{luisroca@um.es}
\affiliation{Departamento de F\'isica, Universidad de Murcia, E-30100 Murcia, Spain.
}

\author{E.~Oset}
\email{Eulogio.Oset@ific.uv.es}
\affiliation{Departamento de F\'isica Te\'orica and IFIC, Centro Mixto Universidad de Valencia-CSIC, 
Institutos de Investigac\'ion de Paterna, 22085, 46071 Valencia, Spain.
}

\begin{abstract}
By looking at the pseudoscalar-vector meson 
spectra in the $\bar{B}\to J/\psi \rho\bK$ 
and $\bar{B}\to J/\psi \bK^*\pi$ weak decays, 
we theoretically investigate the double-pole 
structure of the $K_1(1270)$ resonance by using the 
Chiral Unitary approach to account for the final 
state interactions between the pseudoscalar $(P)$ and 
vector ($V$) mesons. The $K_1(1270)$ resonance is 
dynamically generated through these interactions 
in coupled channels and influences the 
shape of the invariant mass distributions under 
consideration. We show how these 
shapes are affected by the $K_1(1270)$ double-pole 
structure to confront the results from our model 
with future experiments that might investigate 
the $PV$ spectra in these decays.
\end{abstract}



\maketitle

\section{Introduction}
\label{sec:intro}

The observation of the axial vector mesons 
$K_1(1270)$ and $K_1(1400)$ 
\cite{Brandenburg:1975gv,daum} 
were identified as the expected 
$1^+$ strange mesons from the quark model. 
Subsequent experiments have explored their 
properties and decay modes \cite{pdg}. While 
the dominant $K_1(1270)$ decay channel is $\rho K$, 
the $K_1(1400)$ is observed to decay mostly 
through the $K^*(892) \pi$ one. These states have 
usually been studied in terms of the mixing of the strange 
states of the $J^{PC}=1^{++}$ and $J^{PC}=1^{+-}$ nonets (see for example \cite{Suzuki:1993yc,Tayduganov:2011ui,Zhang:2017cbi}).
Other exhaustive analyses of the $1^+$ low-lying 
mesons as dynamically generated resonances 
found that the $S=1$ and $I=1/2$ poles were 
not compatible with the above assignment, but 
rather they should be identified as a double-pole 
structure for the $K_1(1270)$ \cite{Roca:2005nm}. 
The two pole structure for the $K_1(1270)$ resonance 
is not unique: there are several cases where two poles 
were found for hadronic resonances, and a recent 
review can be seen in Ref.~\cite{ulfreview}. 
The discovery of the two-pole structure of 
the $K_1(1270)$ triggered studies looking for 
scenarios where this prediction could be tested. 
The analysis of the $K^-p\to K^-\pi^+\pi^- p$ data 
at $63$ GeV done in Ref.~\cite{daum} provided 
additional support to the existence of two states: 
one with a mass of 1195 MeV coupling mostly to 
the $K^*(892) \pi$ channel, and one with 
a mass of 1284 MeV coupling to the $\rho K$ one. Several 
reactions aimed at observing these two states were proposed, 
such as $D^0 \to \pi^+ VP$ \cite{wang2019}, which is 
similar to $B^- \to  J/\psi  K^-_1(1270)$ with 
the hadronization involving three light mesons. 
More recently, another study considered the 
$D^+ \to  \nu \l^+ K_1(1270)$ decay \cite{wang2020}, 
identifying the signatures in the invariant-mass 
distributions of the decaying $K_1(1270)$. 
On the other hand, expected improvements in the 
experimental capabilities to study $B$-meson 
decays with higher statistics, like in the Belle II 
experiment, make these proposals an interesting 
scenario to look for.

In this work we provide an additional reaction, 
considering decays of the form $\bar{B}^0 \to J/ 
\psi VP$,  where $VP$ are the vector and 
pseudoscalar meson pairs, $\rho\bK$ and $\bK^* 
\pi$, using the chiral unitary approach, and 
we look for signatures of the two $K_1(1270)$ states. 
A related work was done in \cite{wangzhang}, where the reaction 
$B^{-}\rightarrow J/\psi \rho^0 K^{-}$ was studied to look 
for signals of the $Z_c(4000)$; however, simultaneously one 
$K_1(1270)$ showed up in the $\rho\bar K$ mass distribution. 
Here we also look into the $\bar K^* \pi$ channel in order 
to see both $K_1(1270)$ states. 

The work proceeds as follows: In section \ref{sec:form}, 
we present the formalism for the elementary production at 
the quark level, where the different $VP$ channels are 
related by $SU(3)$ arguments. Then, we account for the final 
$VP$ interaction by implementing meson-meson scattering based 
on the Chiral Unitary  approach. In section \ref{sec:res} 
we compute the invariant-mass distribution for the 
$VP$ pair and its structure in terms of the individual 
poles, and identify the regions where the signature 
of such poles can be extracted. 


\section{Formalism}
\label{sec:form}
\subsection{$VP$ pseudoscalar and vector mesons 
production}

\begin{figure}[h!]
\includegraphics[width=0.5\textwidth]{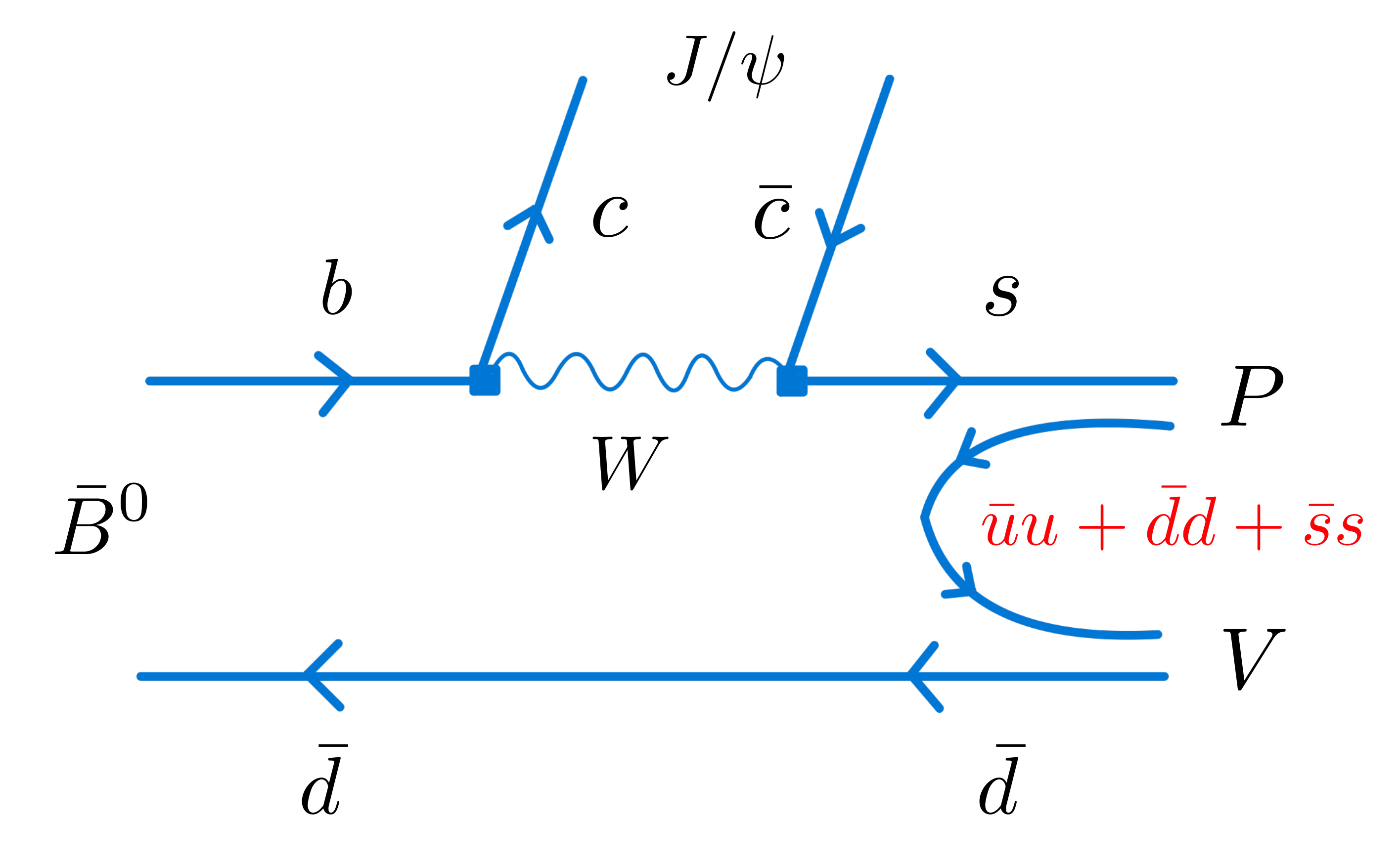}
\caption{Dominant diagram for the $\bar{B} \to 
J/\psi (P\,V)$ reaction at the quark level. 
In the first step, a $b$ quark decays into 
a $c$ one by emission of a gauge boson $W$ which 
then produces a strange quark $s$ along with a 
$\bar{c}$ one. Finally, we are left with a 
$c\bar{c}$ pair, forming the $J/\psi$, and the 
$s\bar{d}$ one. This latter pair is hadronized in order to 
produce a pseudoscalar-vector meson pair emerging as the final state.}\label{Bdecay}
\end{figure}

The relevant contribution for the $\bar{B}\to J/\psi\, VP$ 
reaction is given at the quark level by the diagram
shown in Fig.~\ref{Bdecay}. 
The mechanism starts with a $b\bar{d}$ quark pair, forming 
the initial $\bar{B}^0$ meson, in which the $b$ quark is converted 
into a $c$ quark by emitting a $W$ boson, which then 
produces an anticharm $\bar{c}$ along with a 
strange quark $s$. In the end, we are left with a 
$c\bar{c}$ pair making up a $J/\psi$ meson, considered 
as a spectator, and a $s\bar{d}$ pair. In order 
to produce a pseudoscalar as well as a vector meson, 
a $\bar{q}q$ pair with the quantum numbers of the vacuum is 
added to the already existing $s\bar{d}$ pair, according 
to the $^3P_0$ model \cite{hadro1,hadro2,hadro3}. Therefore, 
the final meson-meson hadronic state has the following 
quark flavor combination:
\begin{equation}
\label{h}
|H\rangle = |s(\bar{u}u+\bar{d}d+\bar{s}s)\bar{d}\rangle \, .
\end{equation}
However, Eq.~\eqref{h} above only refers to the quark 
content of the final mesonic states and it does not carry 
any information about the pseudoscalar or vector nature of 
those hadronic states. This is done by defining the 
$q\bar{q}$-matrix denoted as $M$, written as
\begin{equation}
\label{eq:1}
M=\left(
           \begin{array}{ccc}
             u\bar u & u\bar d  & u\bar s  \\
             d\bar u & d\bar d  & d\bar s  \\
             s\bar u & s\bar d  & s\bar s  
           \end{array}
         \right) \, ,
\end{equation}
in terms of which Eq.~\eqref{h} reads 
\begin{equation}
\label{hstate}
|H\rangle = |s(\bar{u}u+\bar{d}d+\bar{s}s)\bar{d}\rangle 
= \sum_i |s\,\bar{q}_i\, q_i\,\bar{d}\rangle = |M_{3i}\,M_{i2}\rangle 
= |(MM)_{32}\rangle \, .
\end{equation}
The final meson-meson components are found by establishing 
the correspondence between $M$ and the pseudoscalar 
and vector meson $SU(3)$ matrices, that is
\begin{equation}
M \Rightarrow P = \left(
           \begin{array}{ccc}
             \frac{\pi^0}{\sqrt{2}} + \frac{\eta}{\sqrt{3}} + \frac{\eta'}{\sqrt{6}} & \pi^+ & K^+ \\
             \pi^- & -\frac{\pi^0}{\sqrt{2}} + \frac{\eta}{\sqrt{3}} + \frac{\eta'}{\sqrt{6}} & K^0 \\
            K^- & \bar{K}^0 & -\frac{1}{\sqrt{3}}\eta + \sqrt{\frac{2}{3}}\eta' \\
           \end{array}
         \right),
\label{Pmatrix}
\end{equation}
and

\begin{equation}
M \Rightarrow V = \left(
           \begin{array}{ccc}
             \frac{\rho^0}{\sqrt{2}} + \frac{\omega}{\sqrt{2}}  & \rho^+ & K^{*+} \\
             \rho^- & -\frac{\rho^0}{\sqrt{2}} + \frac{\omega}{\sqrt{2}}  & K^{*0} \\
            K^{*-} & \bar{K}^{*0} & \phi \\
           \end{array}
         \right)\, ,
\label{Vmatrix}
\end{equation}
where the standard $\eta-\eta^{\prime}$ \cite{bramon} and $\omega_1-\omega_8$ 
mixings have been used for $P$ and $V$, respectively, in order to 
match the right flavor content of the matrix $M$.

Since we aim at describing a reaction with 
a pseudoscalar along with a vector meson as 
final states, the matrix $M$ in Eqs.~\eqref{Pmatrix} 
and \eqref{Vmatrix} should be combined 
according to Eq.~\eqref{hstate} in such a way 
that it gives the product $PV$ or $VP$. There 
is nothing in our model that privileges one over 
the other, and thus we consider an equal-weighted 
combination between them in Eq.~\eqref{hstate} 
so that it can be rewritten as
\begin{equation}
\label{Hpv}
|H\rangle = |(PV)_{32} \rangle + |(VP)_{32}\rangle \, .
\end{equation}
Therefore, the pseudoscalar and vector mesons 
produced in the 
reaction are
\begin{equation}
\label{Hcb}
|H\rangle=|\,\rho^+\bK^-\,\rangle - \frac{1}{\sqrt{2}}|\,\rho^0\bK^0\,\rangle + 
|\,K^{*-}\pi^+\,\rangle - \frac{1}{\sqrt{2}}|\,\bK^{*0} \pi^0\,\rangle + 
\frac{1}{\sqrt{2}}|\,\omega \bK^0\,\rangle + |\,\phi \bK^0\,\rangle\, \, , 
\end{equation}
where a term corresponding to the $\bK^{*0}\eta$ 
channel has canceled out in the evaluation of 
Eq.~\eqref{Hpv} by using Eqs.~\eqref{Pmatrix} 
and \eqref{Vmatrix}. Note that the procedure 
adopted provides the final $VP$ meson-meson 
components as well as their relative weights, 
which will play a significant role in the 
mass spectrum. 

The final $|H\rangle$ state can be written in 
the isospin basis by considering the following 
multiplets: $(-\rho^+,\rho^0,\rho^-)$, 
$(\bK^0,-K^-)$, $(-\pi^+,\pi^0,\pi^-)$ for the 
$\rho$, $\bK$ and $\pi$ mesons, respectively. 
Then, the final $VP$ states in isospin 
$I=1/2$ are given by
\begin{eqnarray}
|\,\rho\,\bK\,\rangle_{I_3=1/2}^{I=1/2} &=& \sqrt{\frac{2}{3}}\,|\,\rho^+K^-\,\rangle - 
\sqrt{\frac{1}{3}}\,|\,\rho^0\bK^0\,\rangle \, ,\nonumber\\
|\,\bK^*\,\pi\,\rangle_{I_3=1/2}^{I=1/2} &=& -\sqrt{\frac{2}{3}}\,|\,K^{*-}\pi^+\,\rangle + 
\sqrt{\frac{1}{3}}\,|\,\bK^{*0}\pi^0\,\rangle \, .
\end{eqnarray}
Using this, we can recast $|H\rangle$ as
\begin{equation}
|H\rangle = \sqrt{\frac{3}{2}}\,|\,\rho\,\bK\,\rangle_{I_3=1/2}^{I=1/2} - 
\sqrt{\frac{3}{2}}\,|\,\bK^*\,\pi\,\rangle_{I_3=1/2}^{I=1/2} 
+\frac{1}{\sqrt{2}}\,|\,\omega \bK\,\rangle_{I_3=1/2}^{I=1/2} + 
|\,\phi\bK\,\rangle_{I_3=1/2}^{I=1/2} \, .
\end{equation}
This last equation also provides the relative 
weights, denoted as $h_i$, in the 
isospin basis between the $i$th $VP$ channels above. 
They are
\begin{align}
\label{weights}
h_{\rho\bK}&=\sqrt{\frac{3}{2}};  &  h_{\bK^*\pi}&=-\sqrt{\frac{3}{2}}\, ;\nonumber\\
h_{\omega\bK}&=\frac{1}{\sqrt{2}}; &  h_{\phi\bK}&=1\, .
\end{align}

\begin{figure}[h!]
\includegraphics[width=0.35\textwidth]{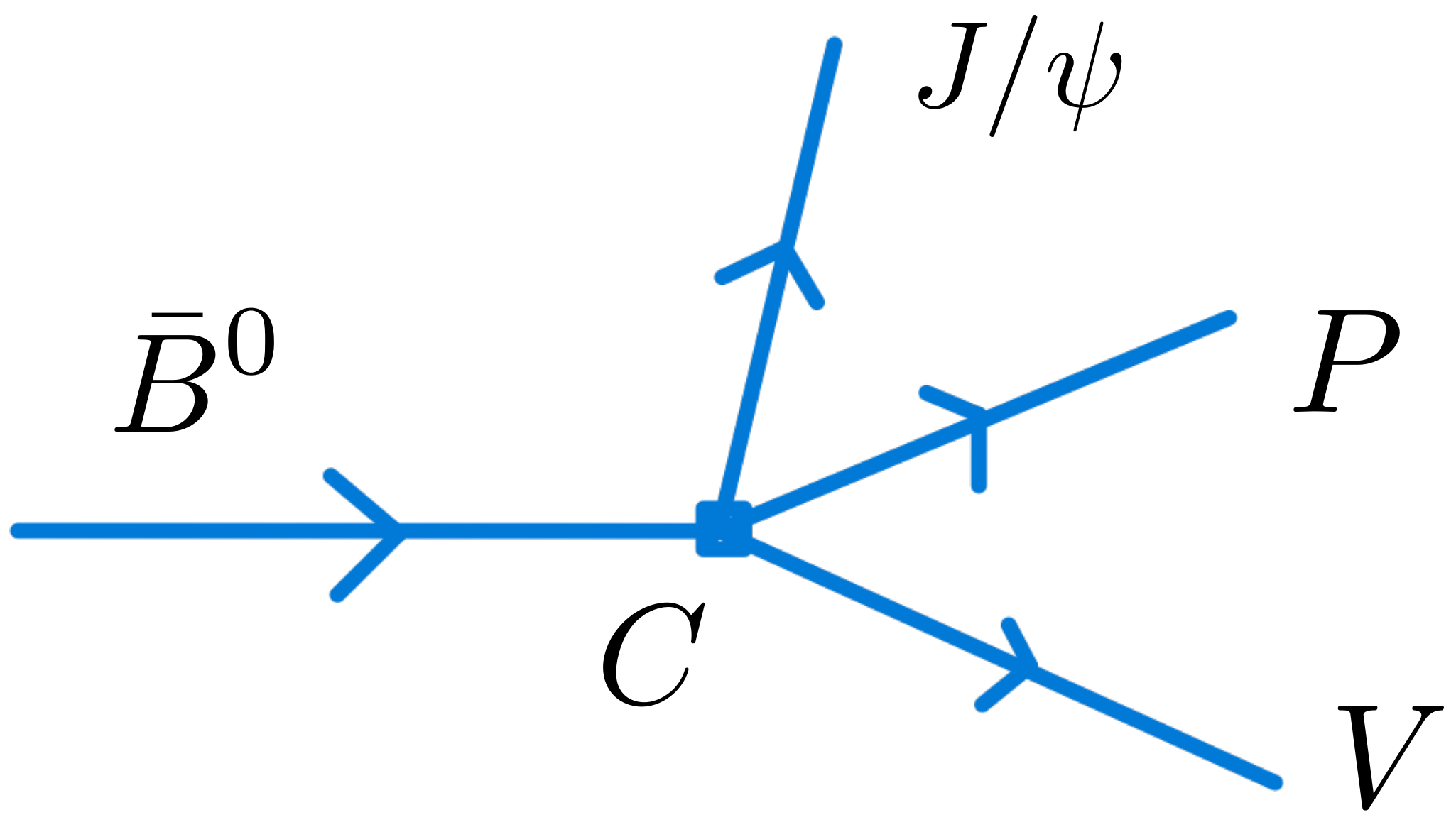}
\caption{Amplitude for the $\bar{B}\to J/\psi (PV)$ decay. The $C$ constant 
is the parametrization of the weak vertex.}\label{btojpv}
\end{figure}

The differential decay width for the $\bar{B}\to J/\psi 
(PV)$ process, illustrated in Fig.~\ref{btojpv}, 
is given by
\begin{equation}
\frac{d\Gamma}{dM_{inv}}=\frac{1}{(2\pi)^3}\frac{1}{4M^2_B}
p_{J/\psi}\tilde{p}_{\pi(\bK)} |T_{B\to J/\psi (PV)}|^2 \, ,
\label{dist}
\end{equation}
where $M_B$ is the $\bar{B}$-meson mass while $p_{J/\psi}$ 
and $\tilde{p}_{\pi(\bK)}$ are the momentum associated with 
$J/\psi$ in the $\bar{B}$ rest frame and $\pi(\bK)$ mesons 
in the $PV$ rest frame, respectively. As a function of 
the $PV$ invariant mass, $M_{inv}$, they are
\begin{equation}
p_{J/\psi}=\frac{\lambda^{1/2}(M^2_B,
M^2_{J/\psi},M^2_{inv})}{2M_B}\, ,
\end{equation}
\begin{equation}
\tilde{p}_{\bK}=\frac{\lambda^{1/2}(M^2_{inv},M^2_{\rho},
m^2_{\bK})}{2M_{inv}}\,\, \textrm{for the}\,\, \rho\bK \,\, \textrm{channel}\,,
\end{equation}
\begin{equation}
\tilde{p}_{\pi}=\frac{\lambda^{1/2}(M^2_{inv},M^2_{\bK^*},
m^2_{\pi})}{2M_{inv}}\,\, \textrm{for the}\,\, \bK^*\pi \,\, \textrm{channel}\,,
\end{equation}
where $\lambda$ stands for the K\"all\'en function. 

\begin{figure}[h!]
\includegraphics[width=0.9\textwidth]{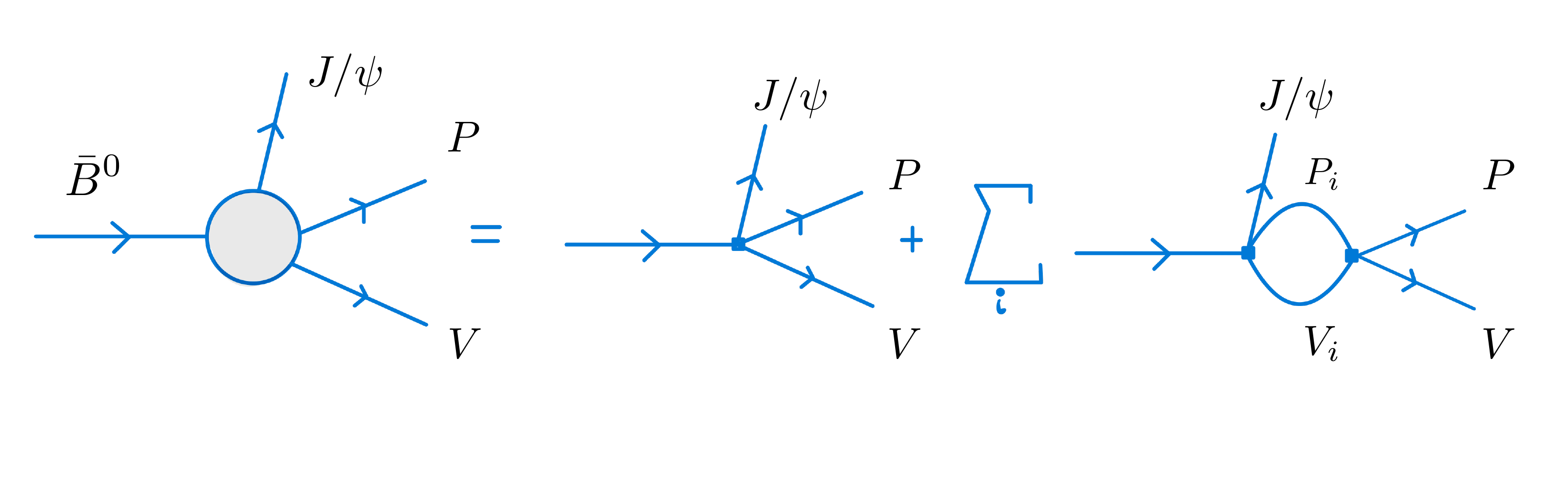}
\caption{Relevant diagrams contributing to the amplitude 
$T_{\bar{B}\to J/\psi (PV)}$ implementing the final state 
interaction. The first diagram on the right-hand side 
corresponds to the tree-level. On the other hand, the second one 
with a loop encodes the final state interaction mechanism and, actually 
it is a sum over all $P_iV_i$ pseudoscalar and vector mesons associated 
with the $i$-channel: $i=1$ for $\phi \bK$, $i=2$ for 
$\omega \bK$, $i=3$ for $\rho\bK$, $i=4$ for $\bK^*\pi$.}\label{diags}
\end{figure}

For the evaluation of the full decay amplitude, 
which is needed in Eq.~\eqref{dist}, we have to consider 
the diagrams in Fig.~\ref{diags}, where 
the final-state interaction mechanism 
is implemented to take into account the $K_1(1270)$ 
resonance contribution for the invariant-mass 
spectra we are interested in. Since we are interested 
in the distributions with $\rho \bK$ and $\bK^*\pi$ as 
final $VP$ states, we have
\begin{equation}
\label{rk}
T_{\bar B\to J/\psi \rho\bK}=C\,(\vec{\epsilon}_{\psi}\cdot
 \vec{\epsilon}_{\rho})
\Big(h_{\rho\bK} +  \sum_i\,h_i
\,G_i(M_{inv})\,t^{I=1/2}_{i\to\rho\bK}(M_{inv})\Big)\, ,
\end{equation} 
and 
\begin{equation}
\label{kpi}
T_{\bar B\to J/\psi \bK^*\pi}=C\,(\vec{\epsilon}_{\psi}\cdot
 \vec{\epsilon}_{\bK^*})
\Big(h_{\bK^*\pi} +  \sum_i\,h_i
\,G_i(M_{inv})\,t^{I=1/2}_{i\to\bK^*\pi}(M_{inv})\Big)\, ,
\end{equation} 
where the index $i$, running from $1$ to 
$4$, stands for each possible $VP$ channel 
involved in the loop, and $C$ contains the information 
of the strength of the weak decay at the tree-level. 
The channels are: $i=1$ for 
$\phi \bK$, $i=2$ for $\omega \bK$, $i=3$ 
for $\rho\bK$, $i=4$ for $\bK^*\pi$. These 
loops are represented by $G_i(M_{inv})$ 
which is the $G$-loop function (given as a 
function of the invariant mass $M_{inv}$) which 
we will discuss in Subsection \ref{fsi}. 
The $h_i$'s are the relative weights in the isospin 
basis, defined in Eq.~\eqref{weights} above. Moreover, 
$\vec{\epsilon}_{\psi}$, $\vec{\epsilon}_{\rho}$, $\vec{\epsilon}_{\bK^*}$ 
are the polarization vectors for the $J/\psi$, 
$\rho$, and $\bK^*$ mesons, respectively.

Furthermore, the amplitudes $t^{I=1/2}_{i\to \rho\bK}$ 
and $t^{I=1/2}_{i\to\bK^*\pi}$ 
in Eqs.~\eqref{rk} and \eqref{kpi} are the 
two-body scattering amplitudes for all possible 
transitions from the $i$th channel to $\rho\bK$ 
(and to $\bK^*\pi$), which in our approach 
encode the resonance $K_1(1270)$ as dynamically 
generated through these interactions \cite{geng2007}, 
and are explained in the following subsection.

\subsection{Final-state interaction and the $K_1(1270)$ resonance}
\label{fsi}

Once the final meson-meson pair is produced at 
tree-level in the $\bar{B}\to J/\psi \,VP$ reaction, 
they undergo final-state interaction from which the 
$K_1(1270)$ resonance emerges dynamically. 
In fact, in Ref.~\cite{geng2007} this resonance was 
dynamically generated through the $s$-wave interaction between 
the pseudoscalar and vector mesons in the $I=1/2$ channel. 
The final-state interaction mechanism is introduced by 
adopting a unitarization procedure using the Bethe-Salpeter 
equation in coupled channels, from which some hadronic states show up 
as poles in the unphysical Riemann sheets of the 
scattering matrices. This approach is a unitary 
extension of Chiral Perturbation theory, called 
the Chiral Unitary approach \cite{oller1,oller2,oller3}, 
which has allowed to describe many hadronic resonances 
as composite states of mesons and/or baryons. In particular, 
in Ref.~\cite{geng2007} the transition amplitudes $t^{I=1/2}_{i\to j}$ 
appearing in Eqs.~\eqref{rk} and \eqref{kpi} 
were unitarized by solving a coupled-channel scattering 
equation in an algebraic form, written in matrix form as 
\begin{equation}\label{bs}
t = (1-V\,G)^{-1}V\, ,
\end{equation}
where in $V_{ij}$, $t_{ij}$ the indices stand for 
the coupled channels: $1$ for $\phi\bK$, 
$2$ for $\omega\bK$, $3$ for $\rho\bK$, $4$ 
for $\bK^*\pi$ and $5$ for $\bK^*\eta$. In 
addition, $V_{ij}$ is the interaction kernel, which 
corresponds to the tree-level amplitudes 
evaluated for all channels we are considering 
in this work by using the chiral Lagrangians 
from Ref.~\cite{Birse} given by
\begin{equation}
\label{lagbirse}
\mathcal{L}_{VVPP}=-\frac{1}{4f^2}\,
\textrm{Tr}\,\Big(\,[V^{\mu},\partial^{\nu}V_{\mu}]\,
[P,\partial_{\nu}P]\,\Big) \, ,
\end{equation}
with $f$ the pion decay constant ($f=93$ MeV) and 
$V,P$ are the matrices given in Eqs.~\eqref{Pmatrix} and 
\eqref{Vmatrix}. Furthermore, 
$G_k(s)$ is the meson-meson loop function associated 
to the $k$th channel, which can be regularized either 
by dimensional or cutoff regularization schemes. In the 
present work, we follow Ref.~\cite{geng2007} which 
employs the former scheme. In this case, the 
$G_k(s)$ loop function is given by 
\begin{eqnarray}
\label{loop}
G_k(\sqrt{s})&=&\frac{1}{16\pi^2}\Big\{a(\mu) + ln\frac{M^2_k}{\mu^2}
+\frac{m^2_k-M^2_k+s}{2s}ln\frac{m^2_k}{M^2_k}
+\frac{q_k}{\sqrt{s}}[ln(s-(M^2_k-m^2_k)+2q_k\sqrt{s})\nonumber\\ 
&+&ln(s+(M^2_k-m^2_k)+2q_k\sqrt{s})-ln(-s+(M^2_k-m^2_k)+2q_k\sqrt{s})\nonumber\\
&-&ln(-s-(M^2_k-m^2_k)+2q_k\sqrt{s})]\Big\} \, ,
\end{eqnarray}
where $M_k$ and $m_k$ stand for the vector and 
pseudoscalar meson masses in the $k$th channel, 
respectively. Moreover, $a(\mu)$ is the subtraction 
constant and in this work we take $a(\mu)=-1.85$ 
for $\mu=900$ MeV, which is the scale of dimensional 
regularization, obtained in Ref.~\cite{geng2007} 
by fitting the experimental $K^-p\to K^-\pi^+\pi^-p$ 
data. In addition, $q_k=|\vec{q}_k|$ is the 
on-shell three-momentum of the meson in the loop, given in the 
center of mass frame by
\begin{equation}
q_k=\frac{\lambda^{1/2}(s,M^2_k,m^2_k)}{2\sqrt{s}}\, .
\end{equation}

The $\rho$ and $\bK^*$ mesons have a relatively 
large width and hence a wide mass distribution. 
In order to take this feature 
into account in our formalism, we convolve the 
loop function $G_k(s)$ with the corresponding 
vector meson spectral function
\begin{equation}\label{sf}
\textrm{Im}[D(s_V)]=\textrm{Im}\Big(\frac{1}{s_V-M_V^2+i M_V\Gamma_V}\Big) \,,
\end{equation}
where $M_V$ stands for the vector meson mass and 
$\Gamma_V$ is the vector meson width, considered 
here as energy independent. Choosing 
an energy-dependent form for $\Gamma_V$, as was done in 
Refs.~\cite{wang2019,wang2020}, does not provide 
any significant change in our results compared with the usual 
uncertainties of our approach. The spectral function 
above is related to the exact propagator for the vector 
meson by using the Lehmann representation, which gives us
\begin{equation}\label{lr}
D(s)=-\frac{1}{\pi}\int\limits_{s_{th}}^{\infty}\, ds_V\,
\frac{\textrm{Im}[D(s_V)]}{s-s_V+i\epsilon}\, ,
\end{equation}
with $s_{th}$ the corresponding vector meson threshold 
for the decay channels with the $\rho$ or $\bK^*$ mesons. 
Therefore, the convolution of the $G_k(s)$ loop function 
defined in Eq.~\eqref{loop} with the vector meson spectral 
function given by Eq.~\eqref{lr} provides
\begin{equation}\label{conv}
G(\sqrt{s},M_k,m_k)=\frac{\int\limits_{(M_V-2\Gamma_V)^2}^{(M_V+2\Gamma_V)^2}\, 
ds_V\, G(\sqrt{s},\sqrt{s_V},m_k)\times \textrm{Im}[D(s_V)]}{
\int\limits_{(M_V-2\Gamma_V)^2}^{(M_V+2\Gamma_V)^2} ds_V\,\textrm{Im}[D(s_V)]}\, ,
\end{equation}
where the limits $(M_V\pm 2\Gamma_V)^2$ are 
considered to be a reasonable cut in the 
integration above. These cuts cause a small deviation 
of the normalization of the Breit-Wigner distribution 
encoded in the spectral function in Eq.~\eqref{sf} 
and in order to reestablish it we divided by the normalization 
integral defined in the denominator in Eq.~\eqref{conv}.

By looking for poles of Eq.~\eqref{bs} in unphysical 
Riemann sheets of the complex $\sqrt{s}$ variable, 
two poles are found in the $I=1/2$ channel, a 
broader one at $\sqrt{s_p}= (1195 - i123)$ 
MeV, and a narrower one at $\sqrt{s_p}= 
(1284 - i73)$ MeV where, for poles not very far from 
the real axis, $\sqrt{s_p}$ can be approximated by 
$\sqrt{s_p}=(M_p-i\Gamma/2)$ in which the real 
part stands for the pole mass, whereas the imaginary 
one is associated with half the width. For the 
sake of convenience, we shall refer to the former 
and latter as pole $A$ and $B$, respectively. 
For the sake of completeness, we show in Table \ref{tab1} the 
parameters obtained in Ref.~\cite{geng2007} for the 
numerical calculation of Eq.~\eqref{bs} described 
in this section, and the couplings to the $i$th-channel 
$g_i^{A(B)}$ of each $K_1(1270)$ pole.

\begin{table}[htp]
\caption{Parameters obtained in Ref.~\cite{geng2007}. In the 
first line: values for the subtraction constant $\alpha$, 
the scale of dimensional regularization $\mu$ along with 
the two $K_1$ poles are given. On the lower part: the channels 
and their corresponding couplings $g_i^{A(B)}$ 
to each $K_1$ pole $A$ and $B$.}
\begin{center}
\begin{tabular}{c|c|c}
\hline\hline
$\alpha(\mu)$~~~~~~~~~~~$\mu$ scale ~~~~& Lower pole & Higher pole\\
 ~~  ~~& (A) & (B)\\
\hline
-1.85~~~~~~~~~~ 900 ~~&~~ (1195 - i123) MeV ~~&~~ (1284 - i73) MeV\\
\hline\hline
 Channels & Couplings $g_i^A$ & $g_i^B$\\ 
 \hline
$\phi\bK$ & $2096-i1208$ & $1166-i774$\\
$\omega\bK$ & $-2046+i821$ & $-1051+i620$\\
$\rho\bK$ & $-1671+i1599$ & $4804+i395$\\
$\bK^*\pi$ & $4747-i2874$ & $769-11171$\\
\hline\hline
\end{tabular}
\end{center}
\label{tab1}
\end{table}%

This will be important in order to study the behavior 
of the distributions given in Eq.~\eqref{dist} 
considering each pole contribution individually. 
The amplitudes given in Eq.~\eqref{bs} contain the 
information about the whole dynamics for the $VP$ 
interaction, including the resonance structure. 
Although both poles are intertwined in the highly 
nonlinear dynamics involved in the amplitude of 
Eq.~\eqref{bs}, it is also interesting for illustrative 
purposes to differentiate the contribution from each 
individual pole. Since it is not possible to directly isolate 
each pole contribution from Eq.~\eqref{bs}, this 
task can be achieved by adopting a Breit-Wigner approach 
for those amplitudes. Then, at the pole position, 
we have
\begin{equation}
\label{bw}
t^{I\,A(B)}_{i,j}=\frac{g^{A(B)}_i\,g^{A(B)}_j}{s-s_p}\, ,
\end{equation} 
where $s_p$ is the pole position of the $K_1$ 
poles $A$ and $B$, whereas $g^{A(B)}_{i(j)}$ 
stands for the coupling of the $i(j)$th channel 
to the pole $A(B)$. We know that the closer to 
the real axis these poles are, the better this 
approximation works. In addition, it is expected 
that experimentally these amplitudes are 
parametrized by a Breit-Wigner form so that, by 
adopting it in our formalism, a comparison between 
our results and those from future experiments 
is more reasonable. Furthermore, this parametrization 
can also be used to encode the double-pole $K_1$ structure 
if we assume that the amplitudes are given by a 
double-Breit-Wigner shape defined as
\begin{equation}
\label{dbw}
T_{i,j}=t^{I\,A}_{i,j}+t^{I\,B}_{i,j}\, ,
\end{equation}
where $t^{I\,A}_{i,j}$ and $t^{I\,B}_{i,j}$ 
are given by Eq.~\eqref{bw} 
for poles $A$ and $B$, respectively. 

It is interesting to mention that the Chiral 
Lagrangian of Eq.~\eqref{lagbirse} can be deduced 
from a more general framework $-$ the local hidden 
gauge approach 
\cite{hidden1,hidden2,hidden4,nagahiro} $-$ 
by exchanging vector mesons. This framework allows us to 
address a related source of interaction based on the exchange 
of pseudoscalar mesons. We address these two issues in 
Appendix \ref{appA} and \ref{appB}, respectively.

Examples of reactions similar to ours, which look 
carefully into the final-state interactions of the mesons 
produced, are the $D^0\to K^0_S\pi^+\pi^-$ reaction 
studied in \cite{robert} and the 
$D^+\to K^-\pi^+\pi^+$ reaction studied in \cite{kubis,nakamura}. 
In \cite{robert} one of the mesons was kept as a spectator, while 
Refs.~\cite{kubis,nakamura} dealt with a three-body interacting 
system. In Ref.~\cite{kubis} the transition amplitude is obtained 
as the sum of amplitudes classified in terms of isospin, 
based upon the dominant modes of weak decay, which are
external and internal emission \cite{chau}. The final 
state interaction was taken into account by means of the 
Omn\`es representation in terms of experimental phase 
shifts. More detailed, and using models for the 
final-state interactions, is the work of \cite{nakamura} 
from which we can draw conclusions concerning our present work.

The first consideration to be made is that while 
undoubtedly these works do a very good job concerning 
the final-state interaction of the meson components, they 
rely upon free parameters, some of which depend directly on 
the reaction studied. For instance, Ref.~\cite{kubis} 
required seven complex parameters that were adjusted to 
the data of the reaction, and in Ref.~\cite{nakamura} the 
number of parameters adjusted to the data was of the order of 
$20$, depending on the options. The use of these formalisms 
in a new reaction would contain unknown parameters. 
Here we benefit from the fact that we consider the $J/\psi$ 
interaction with the light mesons to be weak, as found in the study 
of coupled channels in \cite{raquelxyz}, and hence we only 
have to worry about the vector-pseudoscalar interaction of 
$\rho \bK$ together with $\bK^*\pi$. On the other hand, we 
do not pretend to reproduce the whole phase space of the 
reaction, but rather a narrow region of the $\rho\bK$ and $\bK^*\pi$ 
invariant masses around the peaks of the $K_1(1270)$ resonances 
that we find. For this purpose, the work of \cite{Roca:2005nm} 
for the vector-pseudoscalar interaction, which predicted two 
$K_1(1270)$ resonances that were tested against data 
of the $K^- p\to K^-\pi^+\pi^- p$ reaction of 
Ref.~\cite{daum} in Ref.~\cite{ulfreview}, is sufficiently 
accurate. Also, limiting ourselves to a narrow region we 
do not have to worry about possible contributions from scalar 
and tensor terms, which are considered in \cite{kubis,nakamura}.

At this point, we have to address a problem concerning the 
$VP$ interaction that was not considered in Ref.~\cite{Roca:2005nm}. 
Indeed, in Refs.~\cite{Roca:2005nm,geng2007} the source of $VP$ 
interaction was given by vector exchange extracted from 
the local hidden gauge approach \cite{hidden1,hidden2,hidden4}. 
We prove in Appendix \ref{appA} that the vector exchange 
interaction leads to the contact chiral interaction of 
Ref.~\cite{Birse} used in Refs.~\cite{Roca:2005nm,geng2007}. 
However, there is another source of interaction based on 
the pseudoscalar exchange, as depicted in Fig.~\ref{FigB1} of 
Appendix \ref{appB}. This interaction was considered in 
Refs.~\cite{misha,kubis,nakamura}. The reason not to consider it 
is analogous to a similar source of interaction considered 
in the $VV$ interaction in Refs.~\cite{raquel,gengvec}. Indeed, in 
these works, this new source of interaction was taken into 
account via the box diagram of Fig.~\ref{FigB2} in Appendix 
\ref{appB}. What was found there was that the real part of the new 
potential was negligible, and only the imaginary part, due to 
the large phase space for $\rho\rho\to \pi\pi$ decay, was 
relevant. We take the opportunity to do the equivalent work here, 
and this is done in Appendix \ref{appB}. In Figs. 
\ref{FigB3}, \ref{FigB4}, \ref{Boxeleven} and \ref{Boxtwelve} 
we show new mechanisms contributing to the $VP$ interaction which 
involve pseudoscalar exchange. We find in all cases a very small 
contribution of a few percent relative to the large terms 
coming from vector exchange.

It is interesting to see that this conclusion agrees with the 
observation made in Ref.~\cite{nakamura}, where the two interaction 
mechanisms $-$ vector exchange and pseudoscalar exchange $-$ were explicitly 
considered [see Figs. 4(a) and 2 of \cite{nakamura}, respectively.] 
The authors in Ref.~\cite{nakamura} stated that ``we found that the effect 
of the diagram Fig. 4(a) connected to $(\pi^+\pi^0)^{I=1}_{P}\bK^0$ 
is the most important among the three-body type diagrams that we consider".
The pseudoscalar exchange part  for the $VP$ interaction is 
considered as a Z graph in Fig.~2 of 
Ref.~\cite{nakamura}. If one selects the Z graphs related to the $VP$ interaction, it is found that the vector exchange has a larger impact on the $\chi^2$ than the pseudoscalar exchange
 \cite{nakamurapriv}. One should add that, using Eq.~\eqref{lagbirse},  one finds that the strength of the vector exchange for the I=1/2 interaction that we consider here is twice as large as the one in the I=3/2 case ($\rho^+\bar K^0$) that is produced in the $D^+$ decay in Ref.~\cite{nakamura}, and it is attractive in I=1/2, while it is repulsive for I=3/2. This further magnifies the relevance of vector exchange in our case.

\section{Results}
\label{sec:res}

\begin{figure}[htp]
  \centering
  \subfigure[]{\includegraphics[width=19pc]{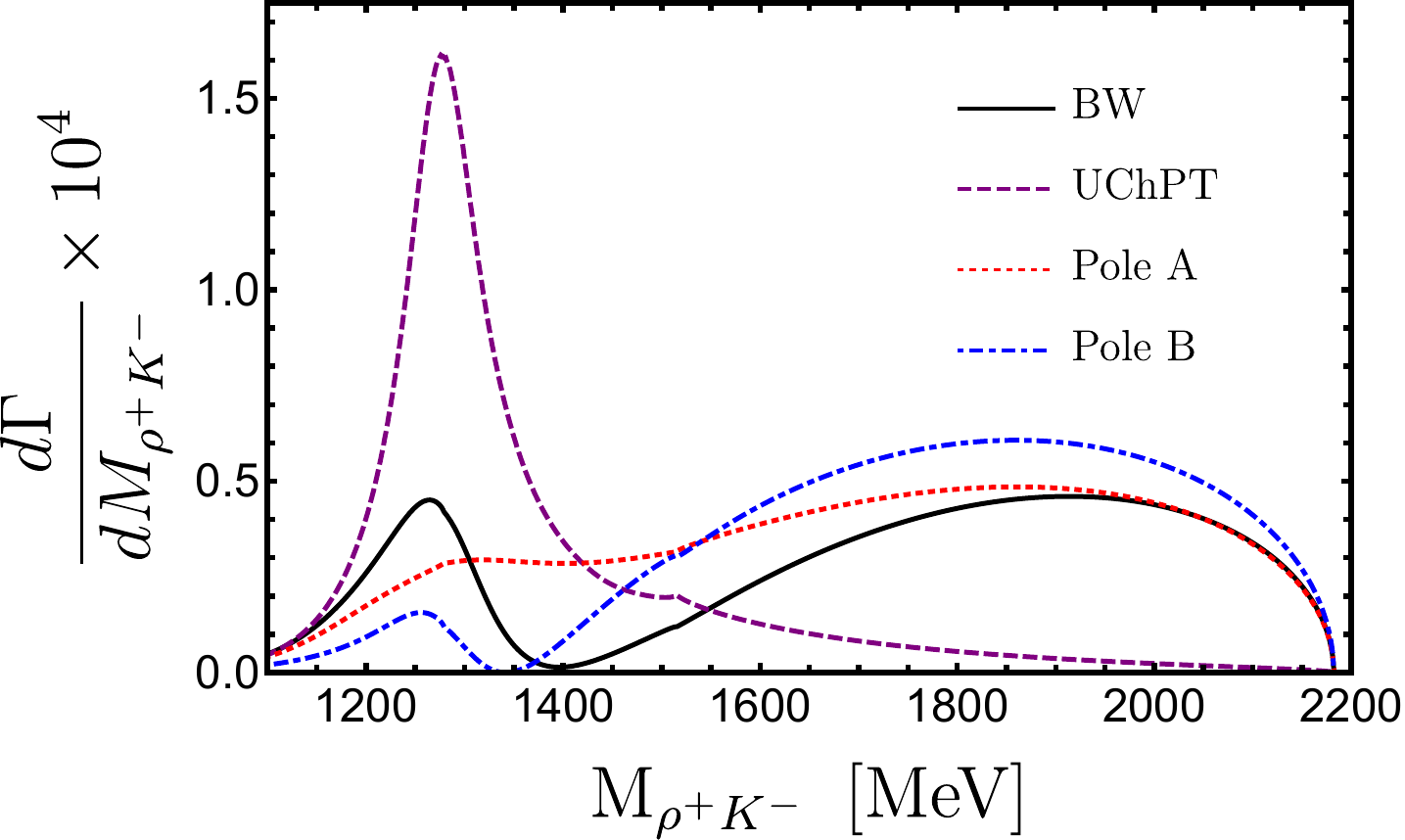}\label{ABrhoK}}\quad
  \subfigure[]{\includegraphics[width=19pc]{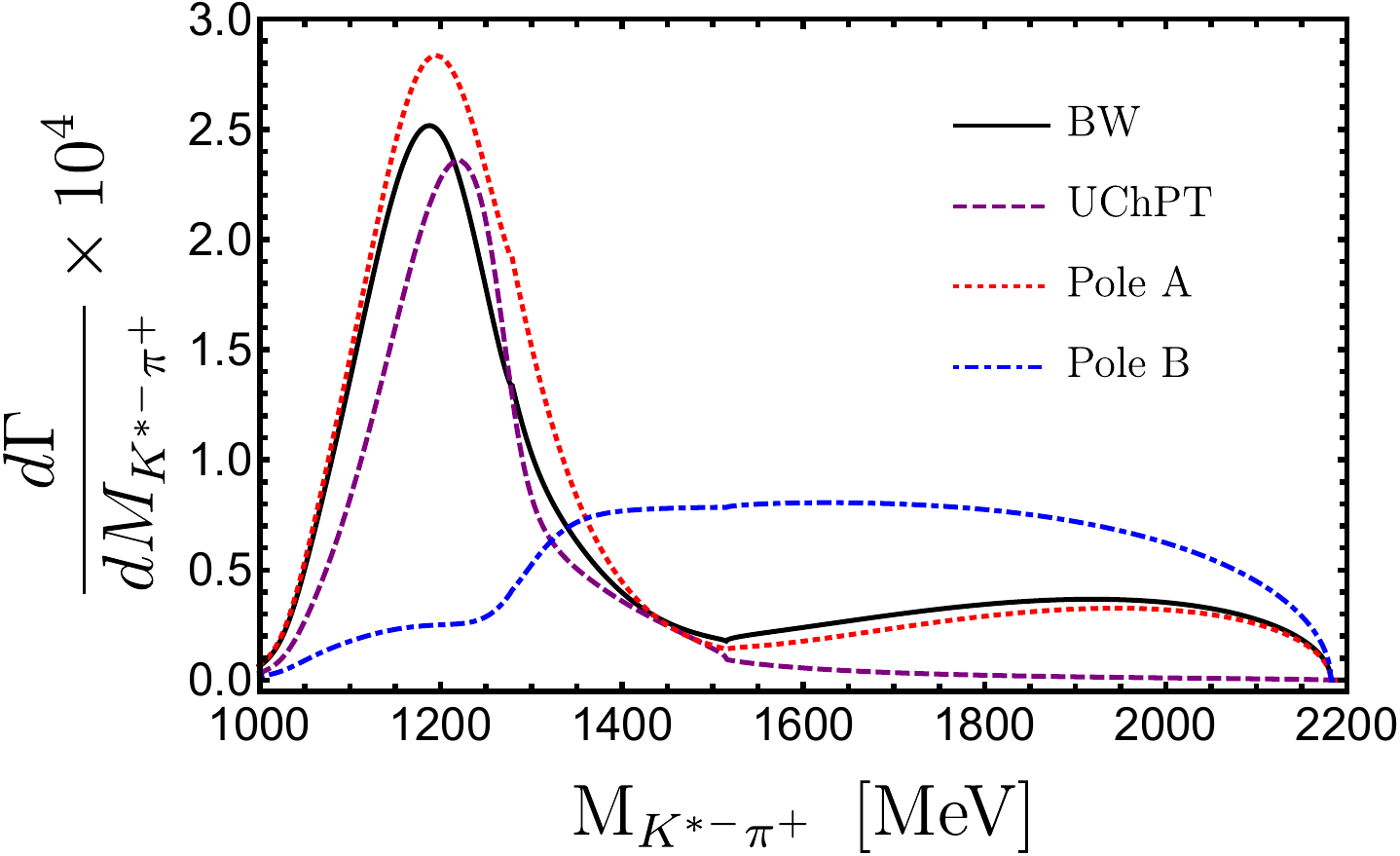}\label{ABKpi}}
  \caption{a) $d\Gamma/M_{\rho^+K^-}$ invariant mass 
distribution for the $\bar{B}^0\to J/\psi\rho^+K^-$ 
reaction (black solid line), compared with the curves 
obtained by considering only: pole $A$ 
(red dotted line), and pole $B$ (blue dot-dashed line); 
b) $d\Gamma/M_{K^{*-}\pi^+}$ distribution for 
$\bar{B}^0\to J/\psi K^{*-}\pi^+$ channel also compared 
with distributions due to each pole contribution 
separately. BW is the Breit-Wigner parametrization whereas UChPT 
stands for Chiral Unitary theory.}
\label{dGammaPoles}
\end{figure} 

In Fig.~\ref{ABrhoK} and \ref{ABKpi} we show the $VP$ 
invariant-mass distributions for $\bar{B}^0\to J/\psi \rho^+ K^-$ 
and $\bar{B}^0\to J/\psi K^{*-}\pi^+$ 
reactions, respectively. The dashed 
lines (labeled UChPT) represent  the results 
obtained using the unitarized amplitudes for the 
two-body $VP$ final-state interaction [Eq.~\eqref{bs}]. 
The solid lines (labeled BW) represent the curves 
obtained if we parametrize the two-body $VP$ amplitudes 
in Eq.~\eqref{bs} by using a double Breit-Wigner-like 
shape [Eq.~\eqref{dbw}]. In Fig.~\ref{dGammaPoles} we 
also show the individual contributions of both poles 
A (dotted line) and B (dot-dashed line) in the Breit-Wigner approach.

Note that the phase space for both 
$\bar{B}^0\to J/\psi \rho^+ K^-$ and 
$\bar{B}^0\to J/\psi K^{*-}\pi^+$ decays take 
nonzero values below the corresponding $VP$ 
threshold as a consequence of the  convolution 
with the vector meson spectral function in order 
to take into account the finite widths of the $\rho$ 
and $\bK^*$ mesons. This effect is especially 
relevant for the $\rho\bK$ channel. 

On the other hand, the  global normalization 
factor in Eqs.~\eqref{rk} and \eqref{kpi} is 
the same for both decay channels, and it does 
not play a relevant role in our results since 
what matters is the relative strengths and shapes
between the mass distribution of the different 
channels and mechanisms considered. Actually, 
this global normalization, $C$, is the 
only free parameter in our model.

It is worth noting that the Chiral Unitary approach 
used in this case has a range of applicability 
up to about $1500-1600$ MeV in the invariant mass. 
We plot the distributions in the whole range, but one 
should bear in mind that the predictions for 
the high invariant masses are less reliable.

A first clear observation from Fig.~\ref{dGammaPoles} 
is that the $K_1(1270)$ resonant shape dominates 
the distributions at low invariant masses. However, each 
distribution is mainly manifesting a different pole 
associated to the $K_1(1270)$. In fact, one would 
expect from the values of the couplings shown in 
Table \ref{tab1} that the pole A would manifest 
more in the $\bar K^*\pi$ distribution and the pole B 
in  $\rho\bK$. Indeed, we see in Fig.~\ref{ABrhoK} 
that the $\rho^+K^-$ mass distribution has a 
pronounced peak at $1284$ MeV, which is just the 
energy region where the highest $K_1(1270)$
pole emerges (pole $B$ in Table 
\ref{tab1}). On the other hand, in Fig.~\ref{ABKpi} 
the $K^{*-}\pi^+$ distribution peaks at $1185$ MeV, 
which is the energy region dominated by the 
lowest $K_1$ pole (pole $A$ in Table \ref{tab1}). 
In addition, the former mass spectrum 
is narrower than the latter, manifesting the 
fact that pole B, which couples mostly to  
$\rho \bar K$, is the narrower one,  with a width 
around $146$ MeV. By contrast, the pole $A$, with a 
width equal to $246$ MeV, is broader than the pole $B$ 
and couples mostly to $\bar K^*\pi$, and then causes the 
peak in the $K^{*-}\pi^+$ distribution to be wider. 

The previous discussion is also applicable 
if we look at the BW curves obtained by using 
the double Breit-Wigner-like amplitudes 
[Eq.~\eqref{dbw}]. The reason of the difference 
between the  UChPT and BW curves is that the 
unitarization amplitudes in Eq.~\eqref{bs} 
contain the full $VP$ dynamics and not just 
the resonant information. We see that this 
difference is more relevant for the $\rho \bar K$ 
distribution. If we look at the individual 
contributions of the different poles, we clearly 
see the dominance of  pole $B$ for the $\rho^+K^-$ 
case and pole $A$ for the  $K^{*-}\pi^+$ case.

\begin{figure}[htp]
  \centering
  \subfigure[]{\includegraphics[width=19pc]{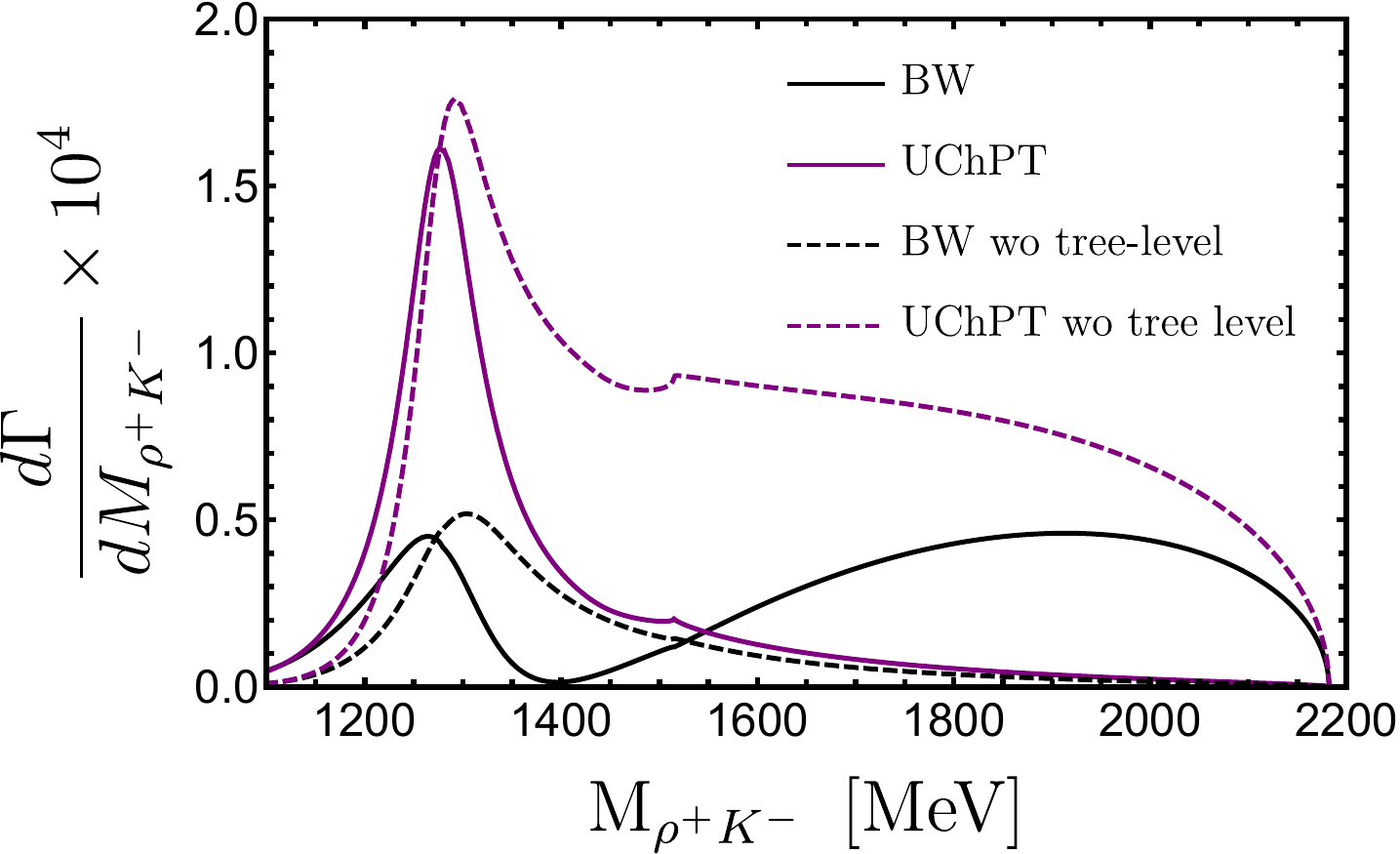}\label{TreerhoK}}\quad
  \subfigure[]{\includegraphics[width=19pc]{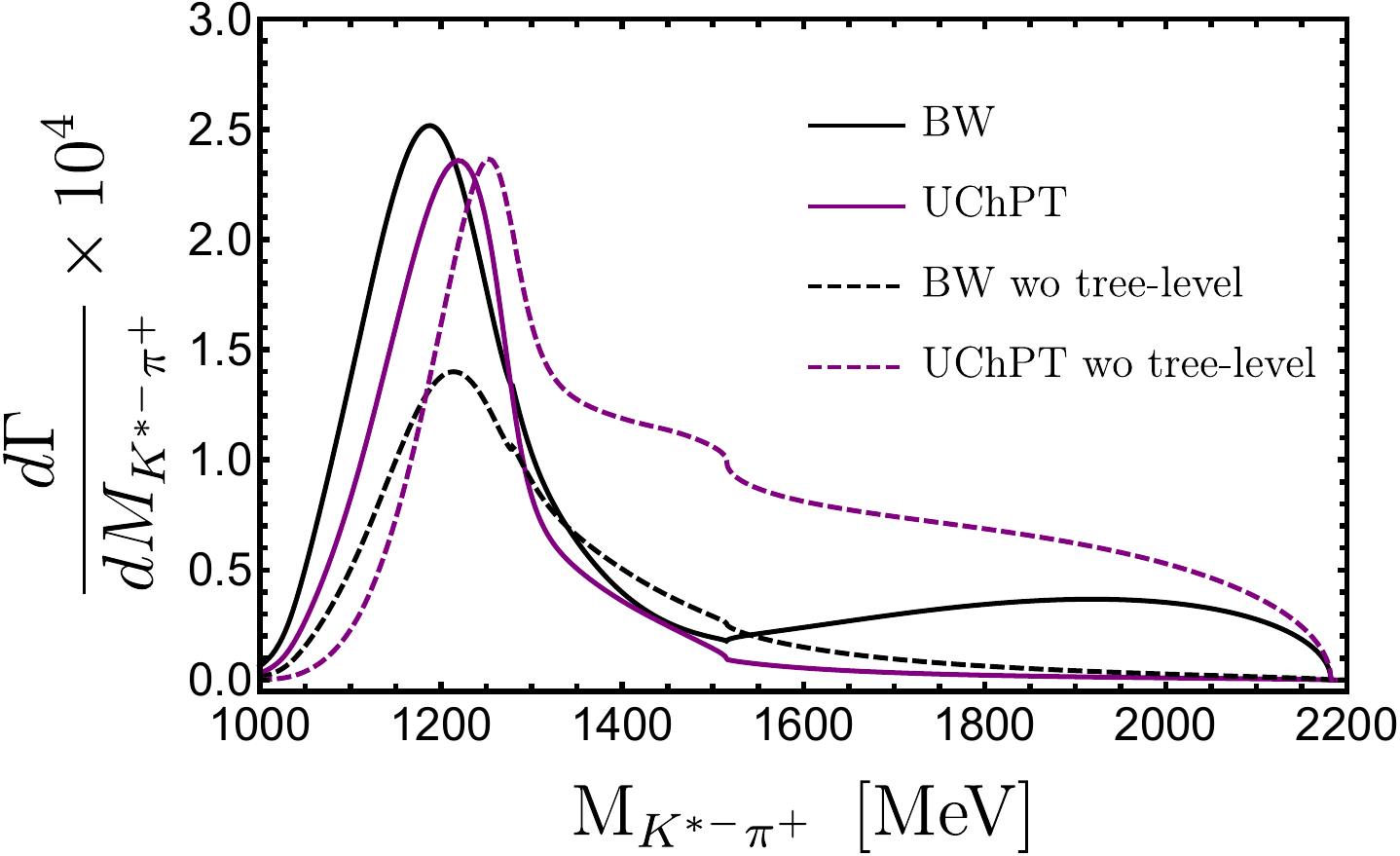}\label{TreerhoK}}
  \caption{a) $d\Gamma/M_{\rho^+K^-}$ invariant mass 
distribution for the $\bar{B}^0\to J/\psi\rho^+K^-$ 
reaction with and without the tree-level mechanism; 
b) $d\Gamma/M_{K^{*-}\pi^+}$ distribution for the 
$\bar{B}^0\to J/\psi K^{*-}\pi^+$ channel also 
compared with the distribution without  
interference between tree-level and the resonant 
part of the amplitude.}\label{dGammaTree}
\end{figure} 

Finally, we study the relative importance 
of the tree-level contribution in Fig.~\ref{diags} 
compared to the final-state interaction (implemented here by using 
the UChPT or the Breit-Wigner approach). 
This is shown in Fig.~\ref{dGammaTree} where 
we confront the results obtained considering 
only the resonant part (dashed lines) with those 
for the whole mechanism: tree-level plus resonant 
parts (solid lines). We can see that for the 
$\bK^*\pi$ channel the shapes of the UChPT curves 
with and without the tree-level contributions 
are similar in strength but shifted by about 
50~MeV. For the double-pole Breit-Wigner 
parametrization the effect of turning off the 
tree-level contribution is more visible in the 
strength of the spectra than in the shift of the 
curves. It decreases the maximum strength of 
the peak by half its value. For the $\rho\bK$ 
channel the UChPT curves exhibit a noticeable 
difference in their shapes at high energies, but 
the strength is not altered much in the resonant 
region.

\section{Conclusions}
\label{sec:conc}

We have theoretically investigated the double-pole 
structure of the $K_1(1270)$ resonance, which was 
shown in Ref.~\cite{geng2007} to be dynamically generated 
through the pseudoscalar-vector meson interaction in 
coupled channels, by looking at the invariant mass 
distributions for the $\rho\bK$ and 
$\bK^*\pi$ pairs, respectively, in the $\bar{B}\to J/\psi \rho\bK$ 
and $\bar{B}\to J/\psi \bK^*\pi$ reactions. The final-state 
interaction mechanism was implemented employing the 
Chiral Unitary approach, in which the pseudoscalar-vector 
meson interaction gives rise to the two $K_1(1270)$ poles 
that, in our model, affect both the $\rho\bK$ and 
$\bK^*\pi$ distributions differently. This feature allows 
us to unveil the double-pole structure in these reactions. 

In particular, we have shown that the $\rho\bK$ 
distribution in the $\bar{B}\to J/\psi \rho\bK$ reaction is 
dominated by the contribution from the $K_1$ highest 
mass pole, whereas the lowest mass pole contributes more for 
the $\bK^*\pi$ distribution in the $\bar{B}\to J/\psi\bK^*\pi$ decay. 
As we have pointed out, this is due to the values of the coupling 
constants of those poles to the different channels considered 
in this work, more specifically, the $\rho\bK$ and $\bK^*\pi$ 
channels. On the other hand, it is important to stress that even 
though it is possible to see one pole dominance over the other in 
each distribution, both $VP$ spectra still have the two $K_1(1270)$ 
poles contributing to their shapes. In view of that, we have also 
modeled the two-body dynamics by using a double-pole Breit-Wigner 
parametrization such that the contributions of the two poles 
could be disentangled. In this case, one expects to observe 
the manifestation of each pole separately in the $VP$ spectra 
to which they couple most strongly. 

An experimental investigation of those 
reactions would be most welcome to shed light on the nature of 
$K_1(1270)$. 

\begin{acknowledgments}

We thank S. X. Nakamura for fruitful discussions.
This work is partly supported by the Spanish 
Ministerio de Economia y Competitividad and 
by Generalitat Valenciana under contract PROMETEO/2020/023, 
and European FEDER funds under Contracts No. 
FIS2017-84038-C2-1-P B and No. FIS2017-84038-C2-2-P B. 
This project has received funding from the European 
Union's Horizon 2020 research and innovation programme 
under grant agreement No. 824093 for the STRONG-2020 project.

\end{acknowledgments}

\appendix
\section{$VP$ interaction in the local hidden gauge approach}
\label{appA}

\begin{figure}[h!]
\includegraphics[width=0.35\textwidth]{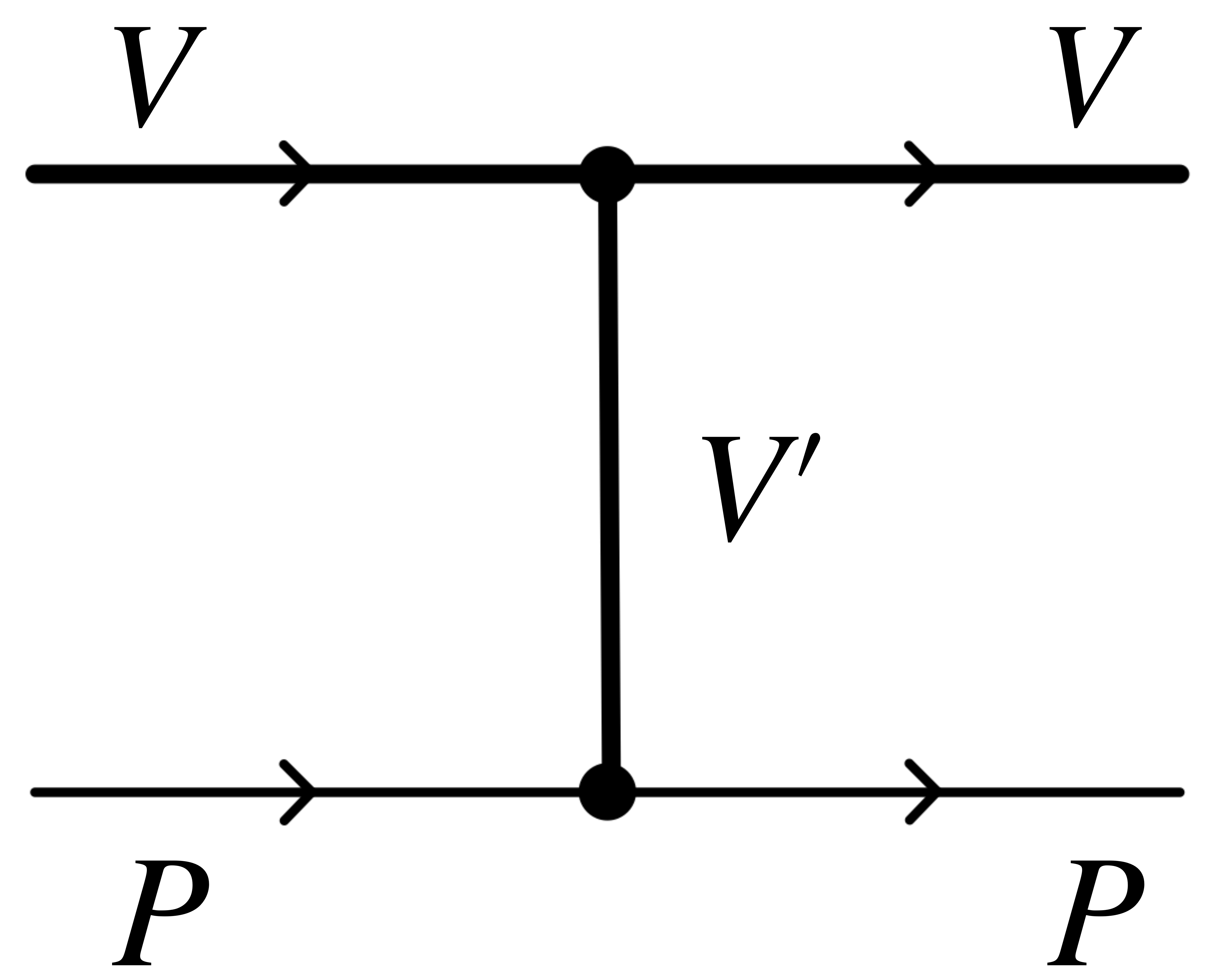}
\caption{$VP$ interaction through the exchange of vector mesons.}\label{FigA1}
\end{figure}

We evaluate the $VP$ interaction in the local 
hidden gauge (LHG) approach
\cite{hidden1,hidden2,hidden4,nagahiro} through 
vector exchange, as depicted in Fig.~\ref{FigA1}.

For this we borrow the $VVV$ and $VPP$ Lagrangians from the LHG, given by
\begin{equation}
\label{lvvv}
\mathcal{L}_{VVV}=i g\langle \,(V_{\mu}\partial_{\nu}V^{\mu}
-\partial_{\nu}V_{\mu}V^{\mu})V^{\nu}\, \rangle \, ,
\end{equation}
where $g=M_V/2f$ ($M_V \approx 800$ MeV, $f=93$ MeV) and 
\begin{equation}
\label{lvpp}
\mathcal{L}_{VPP}=-ig\langle V^{\mu} [P,\partial_{\mu}P]\,\rangle \, ,
\end{equation}
where $\langle \,.\,.\,.\, \rangle$ stands for the trace in $SU(3)$ and 
$V,\,P$ are the matrices given in Eqs.~\eqref{Pmatrix} and \eqref{Vmatrix}. 
The Chiral Lagrangian of Eq.~\eqref{lagbirse} can be obtained from these 
Lagrangians in the following way. First, we make the approximation 
that the three-momenta of the vector mesons are very small with respect 
to the vector meson mass. This is so in our particular case, and hence one 
takes the limit of negligible three-momenta versus the vector meson mass. 
In this case, the vector field $V^{\nu}$ in Eq.~\eqref{lvvv} cannot correspond 
to an external vector of Fig.~\ref{FigA1}. This is so because if it were an external 
vector, then $\nu=1,\, 2,\, 3$ since $\epsilon^0=0$ when $\mathbf{p}_V=0$. But then we 
have $\partial_{\nu}$ which gives rise to a vector three-momentum that is zero. 
Then, $V^{\nu}$ is the $V^{\prime}$ vector exchanged in Fig.~\ref{FigA1} and one has a 
structure like in the $VPP$ Lagrangian, only with the extra factor 
$\epsilon_{\mu}\epsilon^{\prime\mu}=-\boldsymbol{\epsilon}^{\prime}\cdot\boldsymbol{\epsilon}$ for the external vectors.

The amplitude for the diagram of Fig.~\ref{FigA1} is then given as 
\begin{equation}
-it=-g\,(V^{\mu}\partial_{\nu}V_{\mu}-\partial_{\nu}V_{\mu}V^{\mu})_{ij}\,
V^{\nu}_{ji}\, \frac{i}{q^2 - M_V^2}\, V^{\nu\prime}_{lm}\,[P,\partial_{{\nu}^{\prime}}P]_{ml}
\end{equation}
where the indices $i,j,l,m$ are the matrix indices of $P$ and $V$ in 
Eqs.~\eqref{Pmatrix} and \eqref{Vmatrix} written explicitly to obtain the traces.

Since 
\begin{equation}
\sum_{pol} \epsilon^{\nu}_{ji}\,\epsilon^{{\nu}^{\prime}}_{lm} =
 \Big(-g^{\nu{\nu}^{\prime}}+\frac{q^{\nu}q^{{\nu}^{\prime}}}{M^2_V} \Big)\,
 \delta_{jl}\,\delta_{im}\, ,
\end{equation}
we readily obtain, neglecting the term $q^{\nu}q^{{\nu}^{\prime}}/M^2_V$ 
consistently with the approximations done,
\begin{equation}
-it=-i \frac{g^2}{M^2_V} \langle \,(V^{\mu}\partial_{\nu}V_{\mu}-
\partial_{\nu}V_{\mu}V^{\mu})\,[\,P,\partial^{\nu}P\,] \,\,\rangle \, , 
\end{equation} 
and hence
\begin{equation}
\mathcal{L}=-\frac{1}{4f^2}\,\langle \,\, [\,V^{\mu},
\partial_{\nu}V^{\mu}]\,[P,\partial^{\nu}P] \,\rangle \, ,
\end{equation}
which is the Chiral Lagrangian of Ref.~\cite{Birse}, as shown in 
Eq.~\eqref{lagbirse}. This equivalence was already shown in a particular case 
for the $\rho\pi$ interaction in \cite{nagahiro}. 
Here we have made a general derivation.

As shown in Ref.~\cite{Roca:2005nm}, the $s$-wave projected potential 
for transition of channel $i$ to $j$ is given by
\begin{equation}
\label{Vij}
V_{ij}= C_{ij}\,\frac{\boldsymbol{\epsilon}\cdot 
\boldsymbol{\epsilon^{\prime}}}{8 f^2}\, \Big( 3s 
- (M^2 + m^2 + M^{\prime\,2} + m^{\prime\, 2}) -\frac{1}{s}(
M^2 - m^2)(M^{\prime\, 2}- m^{\prime\, 2}) \Big) \, ,
\end{equation}
where $M$, $m$, $M^{\prime}$, $m^{\prime}$ are the initial and 
final vector and pseudoscalar masses, respectively, 
$\boldsymbol{\epsilon}$ and $\boldsymbol{\epsilon^{\prime}}$ the 
polarization vectors of the initial and final vectors, and $C_{ij}$ 
are coefficients given in Table II of Ref.~\cite{Roca:2005nm}. 
Of relevance here are the coefficients 
$C_{\rho K,\rho K}=C_{K^*\pi,K^*\pi}=-2$; $C_{\rho K,K^*\pi}=1/2$. 

\section{Pseudoscalar exchange in the vector pseudoscalar interaction}
\label{appB}

\begin{figure}[h!]
\includegraphics[width=0.49\textwidth]{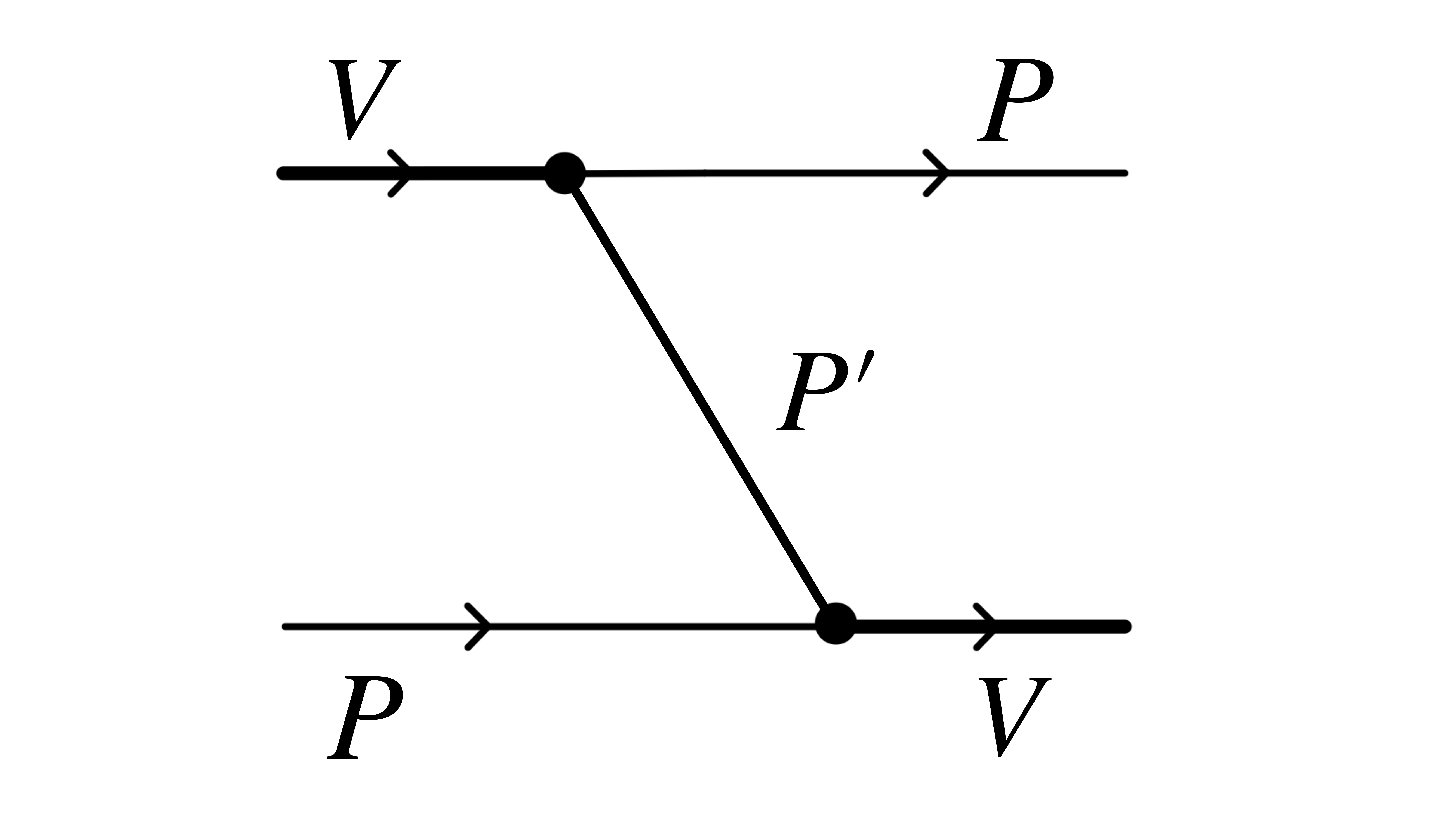}
\caption{Source for $VP$ interaction through the exchange of pseudoscalars.}\label{FigB1}
\end{figure}

In \cite{misha} it was pointed out that a source of the 
$VP$ interaction is given by the diagram in Fig.~\ref{FigB1}. 

This interaction was also considered in \cite{nakamura} in the 
study of the $VP$ interaction in the $D^+\to K^- \pi^+ \pi^+$ reaction; 
however, this was done in addition to the vector exchange discussed here in 
Appendix \ref{appA}, and it was found that the vector exchange is far more important 
than the contribution of pseudoscalar exchange \cite{nakamura,nakamurapriv}. 
We address this issue here in connection with our $VP$ channels 
$\rho \bK$, $\bK^*\pi$ that we have in the problem under study. 

\begin{figure}[h!]
\includegraphics[width=0.49\textwidth]{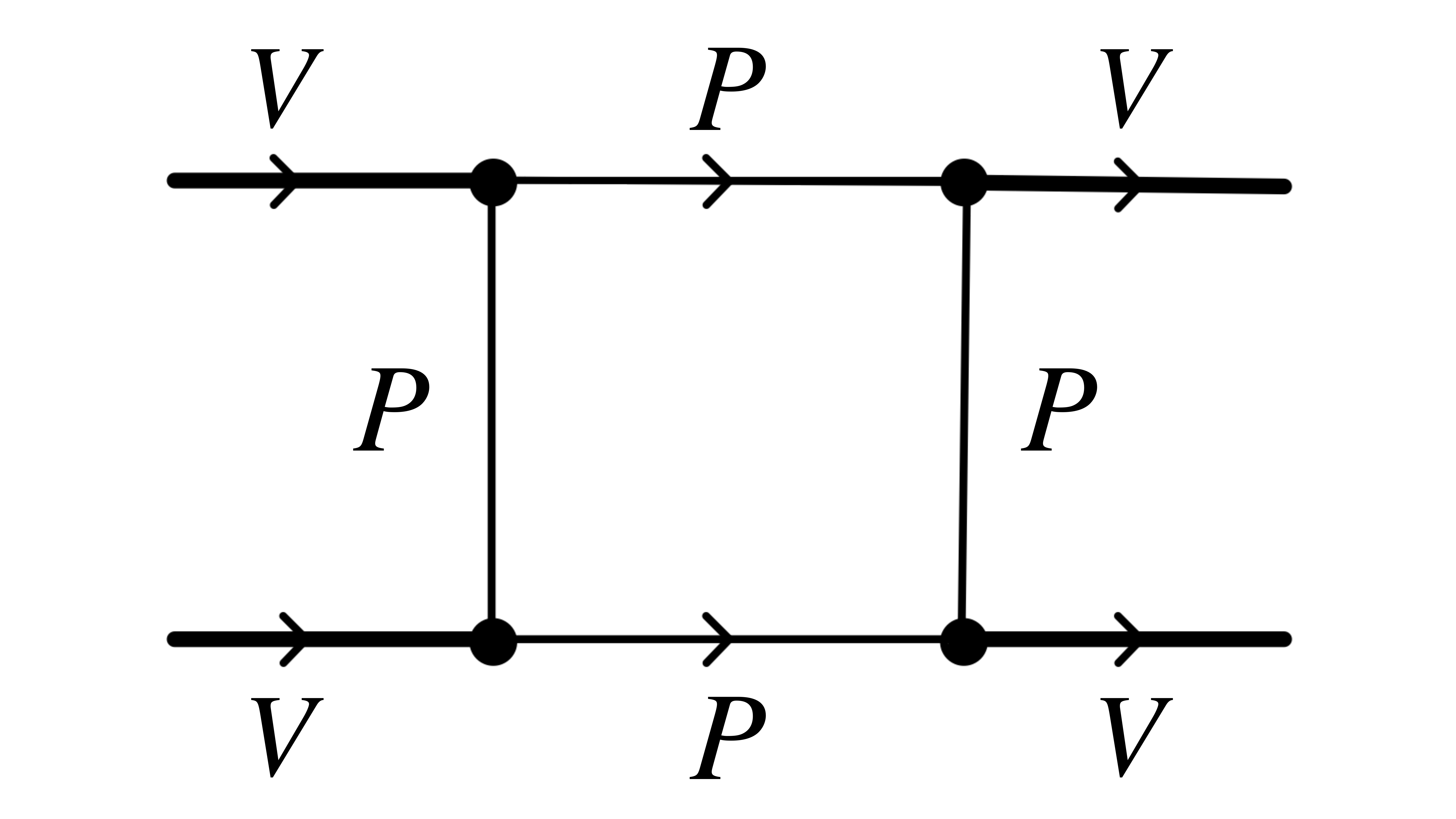}
\caption{Box diagram in the $VV$ interaction given by $P$ exchange and 
intermediate $PP$ states.}\label{FigB2}
\end{figure}

The effect of pseudoscalar exchange was already addressed in the study of the 
vector-vector interaction in \cite{raquel,gengvec}. The box diagram of Fig.~\ref{FigB2} 
was evaluated  exactly with the full structure of the four intermediate 
propagators. Then the $VV\to VV$ potential obtained there was added to the one 
obtained from $VV\to VV$ with a single vector exchanged discussed in Appendix \ref{appA} and 
the whole potential was iterated with the Bethe-Salpeter equation. It was found 
that the real part of the box diagrams was negligible compared to the vector 
exchange, but the imaginary part provided a source of decay for the found $VV$ molecular 
bound states. This was relevant because the bound $VV$ states without this 
term had no width except for a small one when considering the width of the vector 
mesons. However, the $PP$ intermediate states have a small mass and provide a 
large phase space for the decay. The box gave rise to a width of the $VV$ states 
but no change in their mass. 

\begin{figure}[h!]
\includegraphics[width=0.45\textwidth]{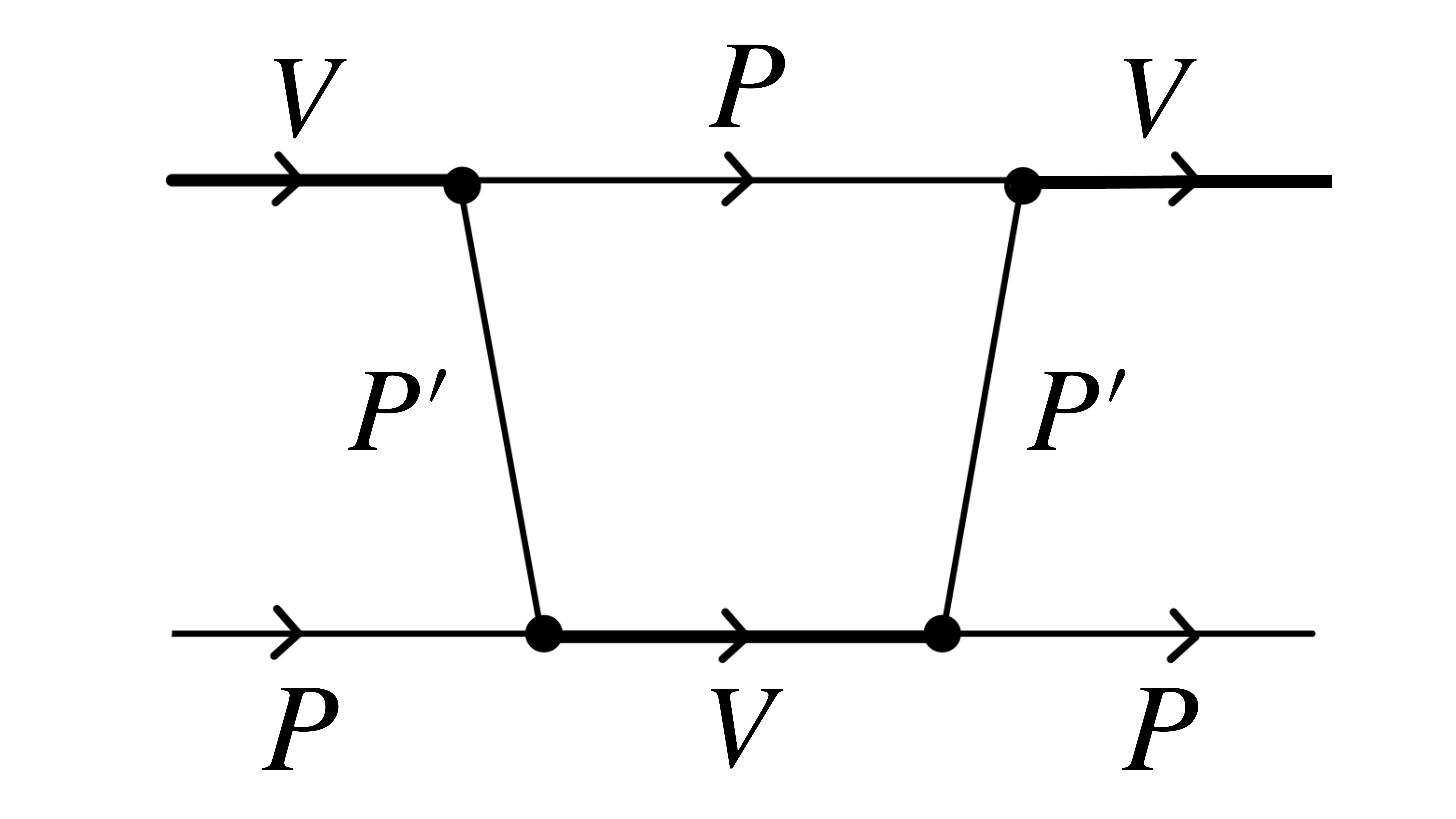}
\caption{Box diagram contributing to $VP\to VP$.}\label{FigB3}
\end{figure}

\begin{figure}[h!]
\includegraphics[width=0.45\textwidth]{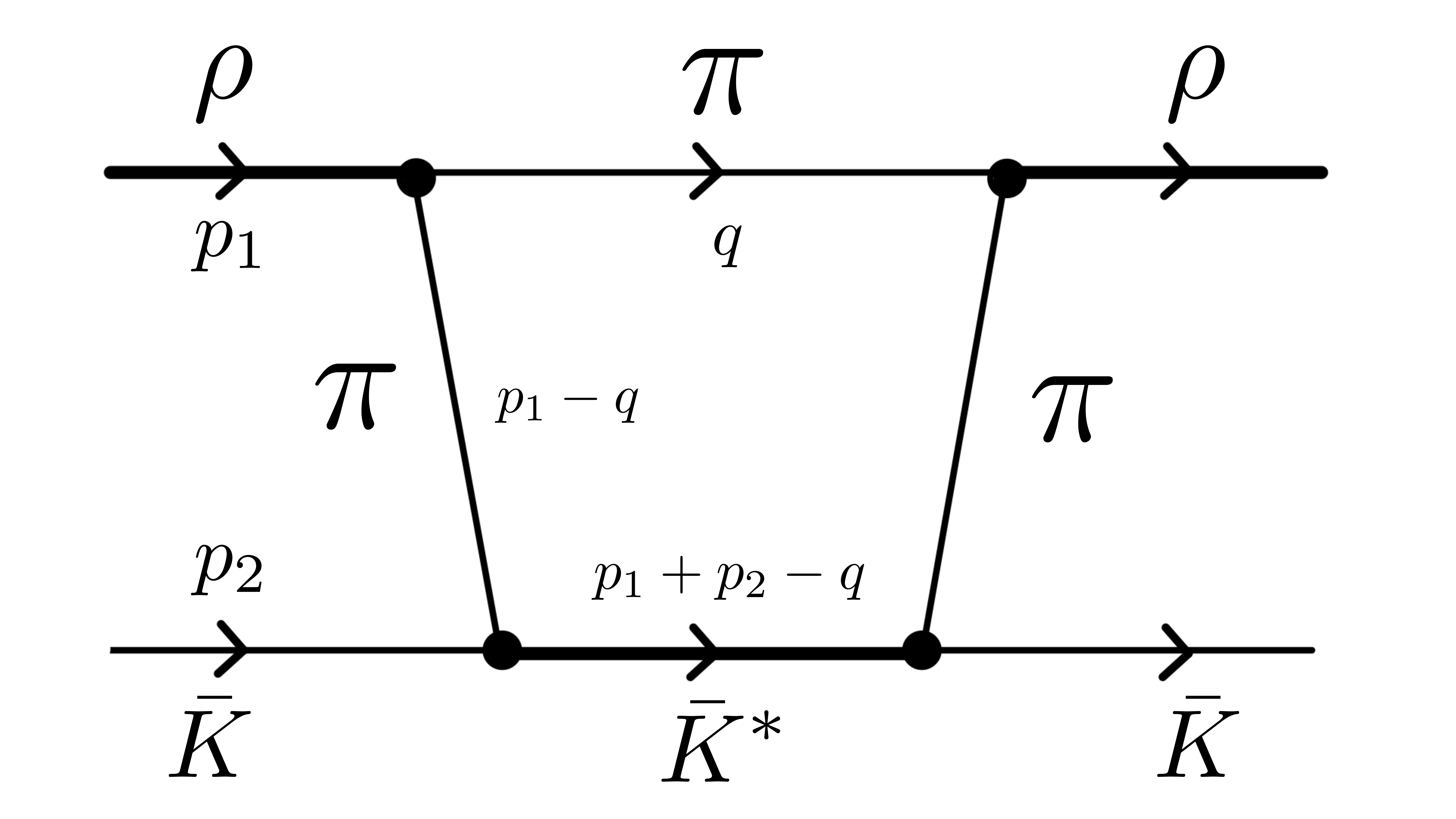}
\caption{Box diagram for $\rho \bar K\to \rho \bar K$ through a $\bK^*\pi$ intermediate state.}\label{FigB4}
\end{figure}

The equivalent diagram to Fig.~\ref{FigB2} for the $VP$ interaction, based on Fig.~\ref{FigB1}, is given in Fig.~\ref{FigB3}.
We evaluate its contribution here for the case of our interest 
$\rho \bK\to \rho \bK$ and the term is depicted in Fig.~\ref{FigB4}.

\begin{figure}[htp]
  \centering
  \subfigure[]{\includegraphics[width=16pc]{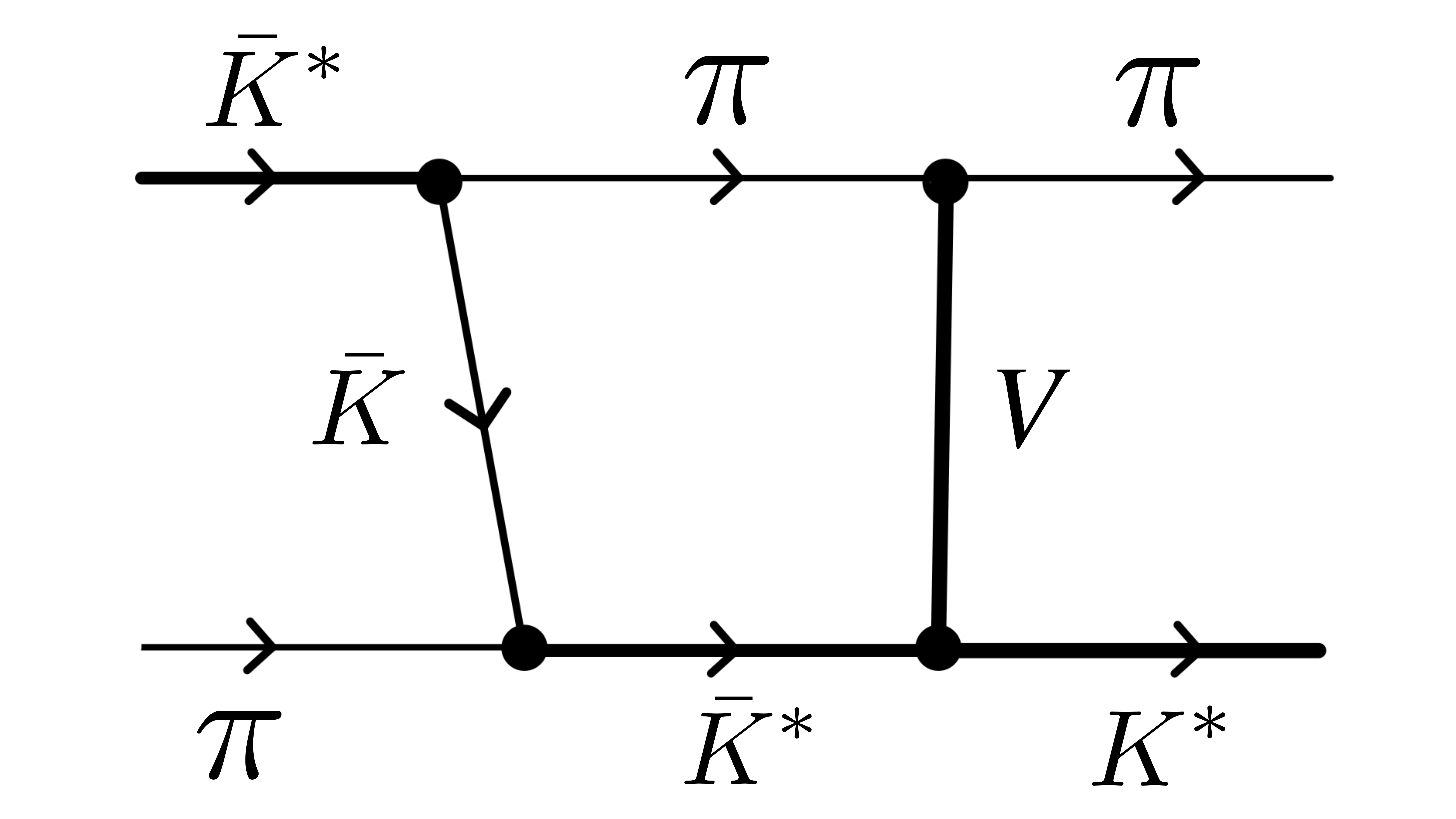}\label{eleven_boxA}}\quad
  \subfigure[]{\includegraphics[width=16pc]{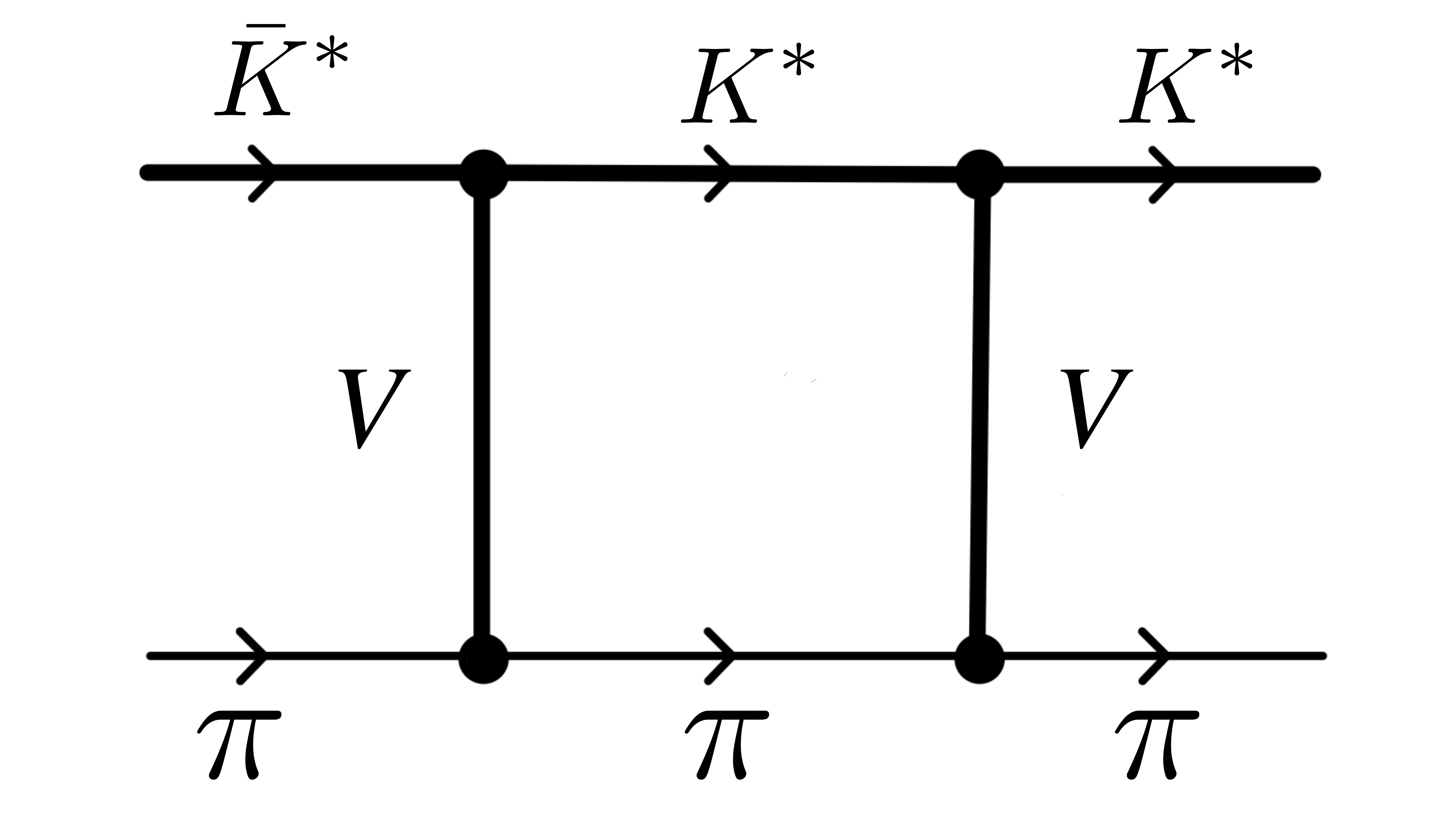}\label{eleven_boxB}}
  \caption{Two steps one meson exchange: a) Pseudoscalar and vector exchange; 
b) Vector and vector exchange.}\label{Boxeleven}
\end{figure}

\begin{figure}[htp]
  \centering
  \subfigure[]{\includegraphics[width=16pc]{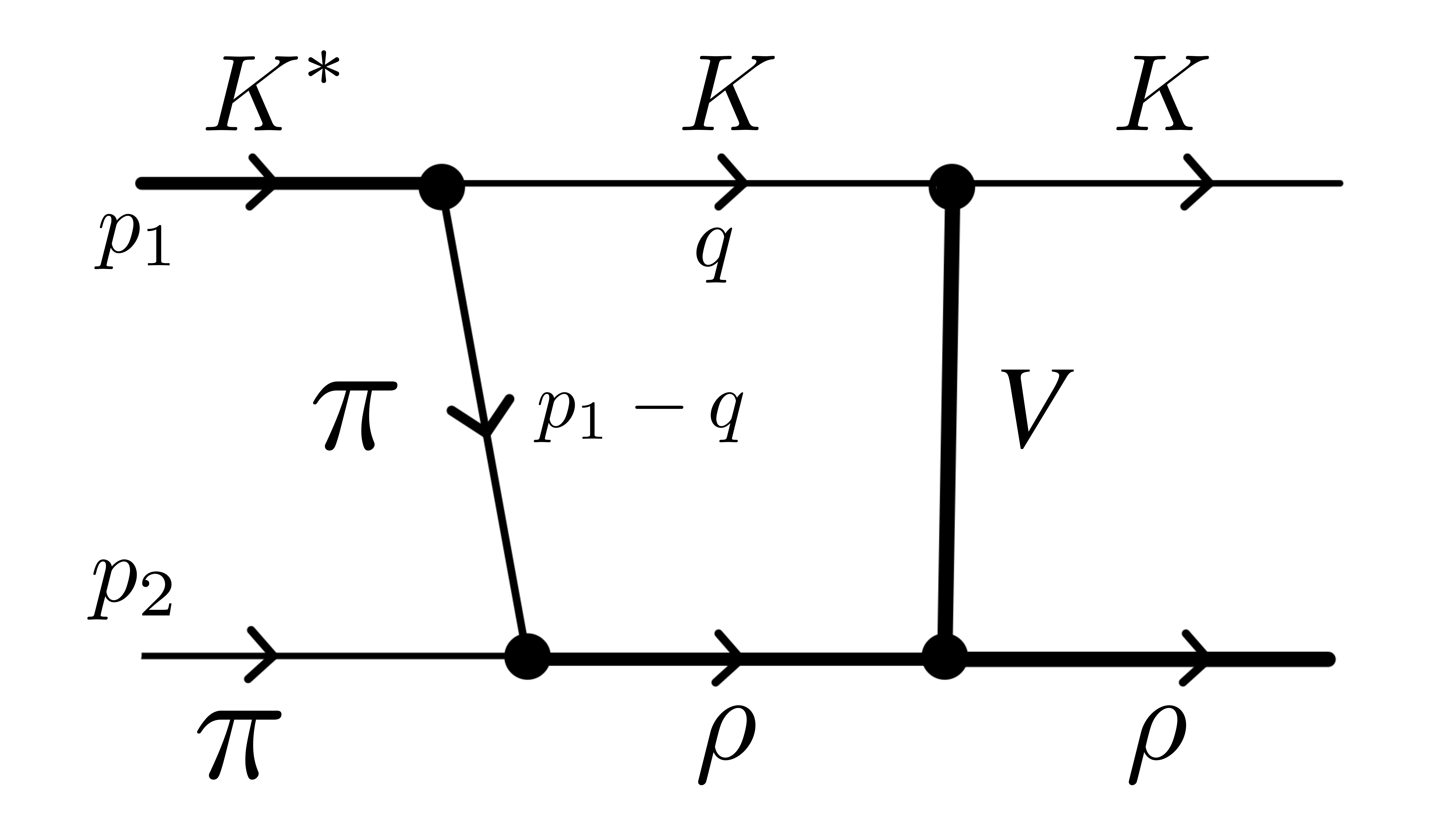}\label{twelve_boxA}}\quad
  \subfigure[]{\includegraphics[width=16pc]{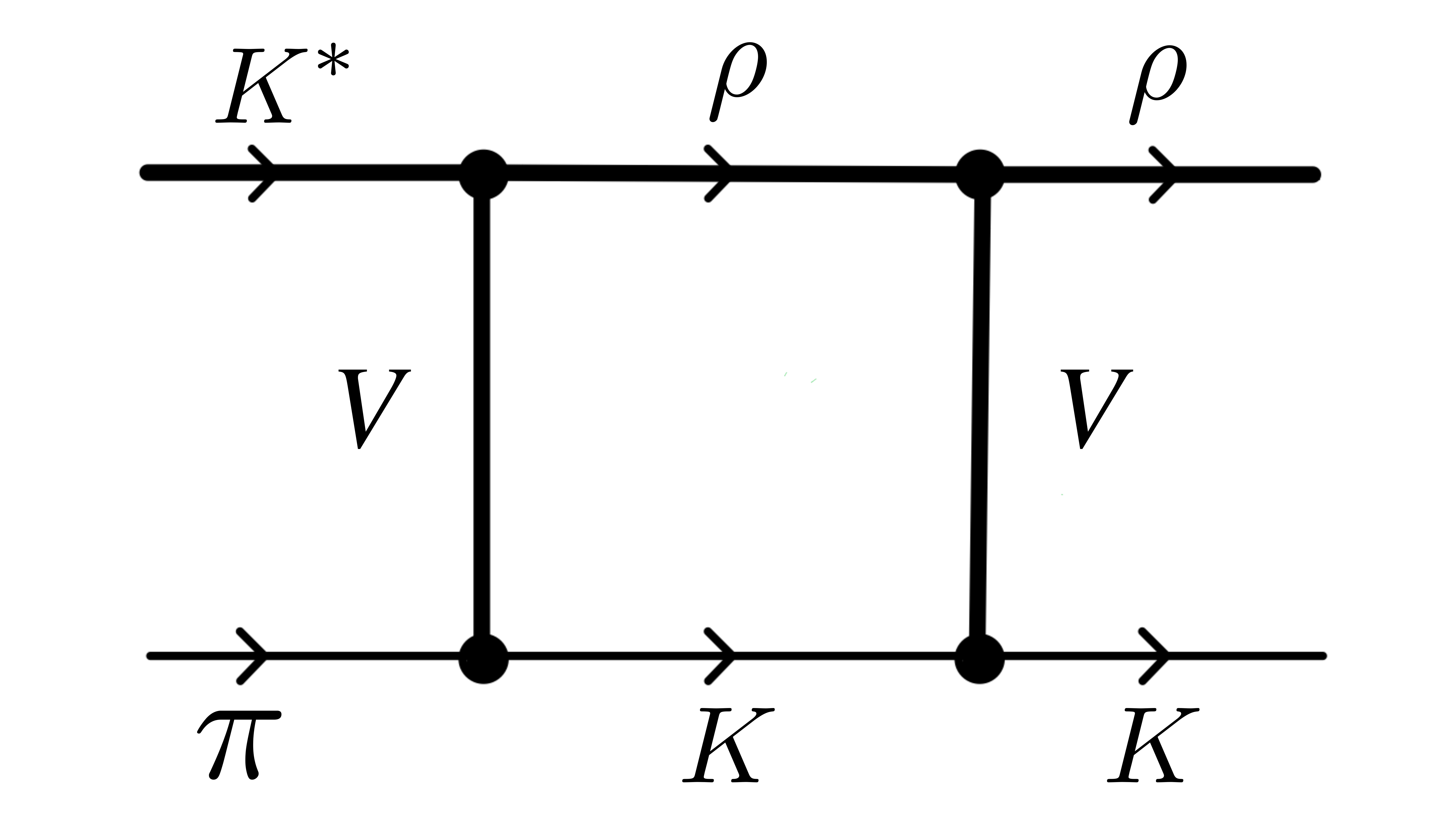}\label{twelve_boxB}}
  \caption{Same as Fig.~\ref{Boxeleven} for the $K^*\pi\to K\rho$ transition.}\label{Boxtwelve}
\end{figure}

However, unlike in the case of $VV$, where we 
need two steps in the $VV\to PP \,\times\, PP \to 
VV$ to obtain a $VV \to VV$ interaction term, here 
a single step $VP\to PV$ driven by pseudoscalar exchange 
already provides a $VP \to PV$ interaction term. 
In order to quantify the relevance of the pseudoscalar 
exchange versus vector exchange, we compare the contributions 
of Fig.~\ref{eleven_boxA} with Fig.~\ref{eleven_boxB} 
and Fig.~\ref{twelve_boxA} with Fig.~\ref{twelve_boxB}. 

We find it sufficient to evaluate the diagrams 
close to the $\rho K$ threshold to benefit from the approximations 
discussed in Appendix \ref{appA} when the three-momenta 
of the vectors compared to their masses are negligible. In this case, we have 
$\epsilon^0=0$ for the $\rho$ and given the fact that the 
large contribution from the diagram involving intermediate 
$K^*\pi$ comes when $\bK^*\pi$ is close to on shell and the $\bK^*$ has a small 
momentum, we also take $\epsilon^0=0$ for the $\bK^*$. In the 
Appendix of Ref.~\cite{ramosakai} it was found that this 
assumption gave surprisingly good results up to relatively 
large momenta of the $\bK^*$ compared to a full 
relativistic calculation for timelike $\bK^*$.

To evaluate the pseudoscalar exchange potential we need the Lagrangian of Eq.~\eqref{lvpp} 
and the isospin structure of the $|\, \rho \bK, I=1/2, I_3=1/2 \,\rangle$ and 
$|\, \bK^*\pi, I=1/2, I_3=1/2 \,\rangle$ states given by
\begin{eqnarray}
|\, \rho \bK, I=1/2, I_3=1/2 \,\rangle &=& \sqrt{\frac{2}{3}} |\rho^+ K^-\rangle - 
\frac{1}{\sqrt{3}}|\rho^0\bK^0\rangle \nonumber \\
|\, \bK^*\pi, I=1/2, I_3=1/2 \,\rangle &=& -\Big(\sqrt{\frac{2}{3}} |\pi^+ K^{*-}\rangle - 
\frac{1}{\sqrt{3}}|\pi^0 \bK^{^*0}\rangle \Big)\, .
\end{eqnarray}
For the vector exchange potentials we use Eq.~\eqref{Vij}.

The contribution of the loop diagram of Fig.~\ref{eleven_boxA}
is given by 
\begin{eqnarray}
\label{diag11a}
-i t_{11a} &=& \int \, \frac{d^4q}{(2\pi)^4}\, g^2 \, 
(\boldsymbol{\epsilon}_{K^*}\cdot \boldsymbol{q})\,
(\boldsymbol{\epsilon^{\prime}}_{K^*}\cdot \boldsymbol{q})\, 
\frac{i}{q^2 - m^2_{\pi} +i\epsilon}\,\frac{i}{(p_1-q)^2-m^2_K +i\epsilon}\times\nonumber\\
&\times&\frac{i}{(p_1+p_2-q)^2-m^2_{K^*}+i\epsilon}\, (-i\tilde{V}_{K^*\pi,K^*\pi})\,
(\boldsymbol{\epsilon^{\prime}}_{K^*}\cdot \boldsymbol{\epsilon^{\prime\prime}}_{K^*})
\,,
\end{eqnarray}
with $\tilde{V}_{ij}$ given by Eq.~\eqref{Vij} removing the 
$\boldsymbol{\epsilon}\cdot \boldsymbol{\epsilon}^{\prime}$ factor. 
The $q^0$ integration is performed analytically using contour integration. For 
this purpose, we use
\begin{equation}
\label{propdecomp}
\frac{1}{q^2-m^2_{\pi}}= \frac{1}{2\omega(q)}\,\Big( 
\frac{1}{q^0-\omega(q)+i\epsilon} - \frac{1}{q^0+\omega(q)-i\epsilon}
 \Big) \, ,
\end{equation}
and the same for the $\bK$ propagator. For the $K^*$ propagator, because 
the heavy $K^*$ mass and the fact that it propagates in the $s$ channel, 
where it is close to on shell, or eventually on shell, it is sufficient to 
take into account only the first term in Eq.~\eqref{propdecomp} $-$ the 
positive-energy part of the propagator. 

We further take into account that $\sum_{pol}\epsilon^{\prime}_{\bK^*\,i}\,\epsilon^{\prime}_{\bK^*\,j}=\delta_{ij}$ and that $\int \,d^3q\,\mathbf{q}_i\,\mathbf{q}_j\,
f(\mathbf{q}^2)=\delta_{ij}\,\int\,d^3q\, \mathbf{q}^2\,f(\mathbf{q}^2)/3$, 
which projects into $s$ wave the operator $q_i\,q_j$, and we find
\begin{eqnarray}
t_{11a}&=&-\frac{1}{3}\,g^2\, (\boldsymbol{\epsilon}_{K^*}\cdot 
\boldsymbol{\epsilon}^{\prime\prime}_{K^*})\,\tilde{V}_{K^*\pi,K^*\pi}\,\int\,\frac{d^3q}{(2\pi)^3}\,\boldsymbol{q}^2\,
\frac{1}{8\,\omega_{\pi}\,\omega_K\,\omega_{K^*}}\,\Big\{ \frac{1}{\sqrt{s}-\omega_{\pi}-\omega_{K^*}+i\frac{\Gamma_{K^*}}{2}}\times\nonumber\\
&\times& \Big( \frac{1}{p^0_1-\omega_{\pi}-\omega_K+i\epsilon}
+ \frac{1}{p^0_2-\omega_K - \omega_{K^*}+ i\frac{\Gamma_{K^*}}{2}} \Big) - \frac{1}{p^0_1+\omega_K+\omega_{\pi}}\times\nonumber\\
&\times&\frac{1}{p^0_2-\omega_K - \omega_{K^*}+i\frac{\Gamma_{K^*}}{2}}\Big\}\Big( \frac{\Lambda^2 - m^2_{\pi}}{\Lambda^2+\boldsymbol{q}^2} \Big)^2 \, ,
\end{eqnarray}
where $\omega_i=\sqrt{m^2_i+\boldsymbol{q}^2}$, we include the $K^*$ width, and we 
have added a form factor for the pseudoscalar exchange. We have taken $\Lambda=1200$ MeV, 
was used in Ref.~\cite{raquel}, and in addition we cut the $\boldsymbol{q}$ integration 
to $\boldsymbol{q}_{max}=900$ MeV, an inherent cutoff in the loop integration associated 
with the use of the chiral potentials in Ref.~\cite{Roca:2005nm}. 

Using the same argumentation, we obtain for the diagram of Fig.~\ref{twelve_boxA}
\begin{eqnarray}
t_{12a}&=&\frac{4}{3}\,g^2\,\frac{1}{3}\,(\boldsymbol{\epsilon}_{K^*}\cdot 
\boldsymbol{\epsilon}_{\rho} )\, \tilde{V}_{\rho K,\rho K} \, \int\,\frac{d^3q}{(2\pi)^3}\,
\boldsymbol{q}^2\,\frac{1}{8\,\omega_{\pi}\,\omega_{K}\,\omega_{\rho}}\,\Big\{ 
\frac{1}{\sqrt{s}-\omega_K - \omega_{\rho}+i\frac{\Gamma_{\rho}}{2}}\times\nonumber\\
&\times&\Big(\, 
\frac{1}{p^0_1-\omega_{\pi}-\omega_K +i\epsilon} + \frac{1}{p^0_2-\omega_{\pi}-\omega_{\rho}+i\frac{\Gamma_{\rho}}{2}}\,\Big) - \frac{1}{p^0_1+\omega_{\pi}+\omega_K}\times\nonumber\\
&\times&\frac{1}{p^0_2-\omega_{\pi}-\omega_{\rho}+i\frac{\Gamma_{\rho}}{2}}\Big\}\Big( \frac{\Lambda^2 - m^2_{\pi}}{\Lambda^2+\boldsymbol{q}^2} \Big)^2 \, .
\end{eqnarray}

The diagrams of Figs.~\ref{eleven_boxB} and \ref{twelve_boxB} are easily 
evaluated and we obtain
\begin{eqnarray}
t_{11b} &=& (\boldsymbol{\epsilon}_{K^*}\cdot\boldsymbol{\epsilon}^{\prime\prime}_{K^*})\,
\int\,\frac{d^3q}{(2\pi)^3}\,\frac{1}{2\,\omega_{\pi}}\,\frac{1}{2\,\omega_{K^*}}\,
\frac{1}{\sqrt{s}-\omega_{\pi}-\omega_{K^*}+i\frac{\Gamma_{K}^*}{2}}\,(\tilde{V}_{K^*\pi,K^*\pi})^2\, , \\
t_{12b} &=& (\boldsymbol{\epsilon}_{K^*}\cdot\boldsymbol{\epsilon}_{\rho})\,\int\,\frac{d^3q}{(2\pi)^3}\,\frac{1}{2\,\omega_{K}}\,\frac{1}{2\,\omega_{\rho}}\,
\frac{1}{\sqrt{s}-\omega_{K}-\omega_{\rho}+i\frac{\Gamma_{\rho}}{2}}\,
(\tilde{V}_{K^*\pi,\rho K}\, \tilde{V}_{\rho K, \rho K})\, .
\end{eqnarray}

Next we evaluate the diagram of Fig.~\ref{FigB4} which contains 
two pseudoscalar exchanges. Then, we find for the product of the 
two left vertices of Fig.~\ref{FigB4}, 
\begin{equation}
-i\tilde{t}_{\rho \bK\to \bK^*\pi}= 4 g^2\, (\mathbf{q}\cdot \boldsymbol{\epsilon}_{\rho})\,
(\mathbf{q}\cdot \boldsymbol{\epsilon}_{\bK^*})\, , 
\end{equation}
and for the loop of Fig.~\ref{FigB4} including the propagators we find
\begin{eqnarray}
-it_{\textrm{10}}&=& 16 g^4 \,\int \frac{d^4q}{(2\pi)^4} (\mathbf{q}\cdot 
\boldsymbol{\epsilon}_{\rho})\, (\mathbf{q}\cdot \boldsymbol{\epsilon}_{\bK^*})\, 
(\mathbf{q}\cdot \boldsymbol{\epsilon}_{\bK^*})\, 
(\mathbf{q}\cdot \boldsymbol{\epsilon}^{\prime}_{\rho})\, \Big[\frac{i}{(p_1-q)^2-m^2_{\pi}+i\epsilon}\Big]^2 
\times\nonumber\\
&\times & \frac{i}{q^2-m^2_{\pi}+i\epsilon}\,\frac{i}{(p_1+p_2-q)^2-m^2_{\bK^*}+i\epsilon}\, ,
\end{eqnarray}
which, upon sum over the $\bK^*$ polarizations, 
$\sum_{pol}\epsilon^i_{\bK^*}\,\epsilon^j_{\bK^*}=\delta^{ij}$, can be written as
\begin{eqnarray}
-it_{10}&=&-i\,16\,g^4\,\frac{\partial}{\partial m_{\pi}^{\prime\,2}}\,\Big\{i\,\int \frac{d^4q}{(2\pi)^4}\,
|\mathbf{q}|^2\,(\mathbf{q}\cdot \boldsymbol{\epsilon}_{\rho})\,(\mathbf{q}\cdot \boldsymbol{\epsilon}^{\prime}_{\rho})\,
\frac{1}{(p_1-q)^2-m_{\pi}^{\prime\,2}+i\epsilon}\,\nonumber\\
&\times&\frac{1}{q^2-m^2_{\pi}+i\epsilon}\,
\frac{1}{(p_1+p_2-q)^2-m^2_{\bK^*}+i\epsilon}\,\Big\}\Big|_{m_{\pi}^{\prime}=m_{\pi}} \, .
\end{eqnarray}

The use of the partial derivative with respect to $m_{\pi}^{\prime\,2}$ saves us one propagator, and then by decomposing the propagators as in Eq.~\eqref{propdecomp} 
and keeping only the positive-energy part for the heavy $\bK^*$, we can immediately perform the $q^0$ integration analytically using Cauchy's residues, with the result
\begin{eqnarray}
\label{tbox}
t_{10}&=&\frac{16}{3}\,g^4\,(\boldsymbol{\epsilon}_{\rho}\cdot\boldsymbol{\epsilon}^{\,\,\prime}_{\rho}) \, 
\frac{{\partial}}{\partial m^{\prime\, 2}_{\pi}}\, \int\, \frac{dq\, q^6}{2\pi^2}\,\frac{1}{2\omega}\,
\frac{1}{2\omega^{\prime}}\,\frac{1}{2\omega^*}\,\Big( \frac{\Lambda^2 - 
m^2_{\pi}}{\Lambda^2+\mathbf{q}^2} \Big)^4\, \Big\{ \frac{1}{\sqrt{s}-\omega - \omega^*\,
+\frac{i\Gamma^*}{2}}\times\,\nonumber\\
&\times&\Big( \frac{1}{p^0_1-\omega - \omega^{\prime} +i\epsilon}\, 
+\frac{1}{p^0_2 - \omega^{\prime} - \omega^*+\frac{i\Gamma^*}{2}} \Big)\,
- \frac{1}{p^0_1+\omega+\omega^{\prime}}\,\frac{1}{p^0_2-\omega^{\prime}-\omega^*+\frac{i\Gamma^*}{2}}
 \Big\}\,,\nonumber\\
\end{eqnarray}
where $\omega=\sqrt{m^2_{\pi}+\mathbf{q}^2}$, 
$\omega^{\prime}=\sqrt{m^{\prime 2}_{\pi}+\mathbf{q}^2}$, 
$\omega^*=\sqrt{m^2_{\bK^*}+\mathbf{q}^2}$, $\Gamma^*$ is the $\bK^*$ width, 
and the energy of $VP$ in the rest frame $\sqrt{s}=p^0_1+p^0_2$. 

The magnitude of $t_{10}$ should be compared to the term coming from vector exchange 
$t_{vex}$, which is evaluated following Appendix \ref{appA} and gives
\begin{equation}
t_{vex}=-2\,\frac{1}{4f^2}\,(p_1+p_3)(p_2+p_4)
\,(\boldsymbol{\epsilon}_{\rho}\cdot\boldsymbol{\epsilon}^{\prime}_{\rho})\, ,
\end{equation}
which can be cast as in Eq.~\eqref{Vij} projected in $s$ wave.
We find at the $\rho\bK$ threshold
\begin{eqnarray}
t_{vex}&=&87.78\,; \nonumber\\
t_{10}&=& 0.25-i\,4.10 \, .
\label{eq:tvexbox}
\end{eqnarray}

We summarize the results obtained in Table \ref{tab2} 
for the energy $\sqrt{s}=1270$ MeV, 
which is the nominal energy of the $K_1(1270)$. As discussed above, 
we take the external three-momenta to be zero and the on-shell energies 
$p^0_1$, $p^0_2$ to be $\sqrt{s}=1270$ MeV. For 
$t_{10}$ we use the $\rho K$ threshold dynamics, which leads to 
Eq.~\eqref{eq:tvexbox}.

\begin{table}[htp]
\caption{Values of the different box diagrams calculated at 
$\sqrt{s}=1270$ MeV. The different $t_i$ refer to the corresponding diagram 
of Fig. i.}
\begin{center}
\begin{tabular}{c c c c c}
\hline\hline
& $\tilde{V}_{\bK^*\pi,\bK^*\pi}$ & $\tilde{V}_{\rho K,\rho K}$ & $\tilde{V}_{\bK^*\pi,\rho K}$ & \\
\hline
& $-82.25$ & $-89.34$ & $21.85$ & \\
\hline\hline
$t_{11a}$ & $t_{11b}$ & $t_{12a}$ & $t_{12b}$ & $t_{10}$\\
\hline
$-0.80 - i1.87$ ~~~~&~~~~ $-64.7-i61.1$ ~~~~&~~~~ $-1.96-i3.01$ ~~~~&~~~~ $18.73+i6.73$ ~~~~&~~~~ $0.25-i4.10$\\
\hline\hline
\end{tabular}
\end{center}
\label{tab2}
\end{table}%

As we can see, the contribution of the $t_{11a}$, with $K$ exchange, is 
very small compared to the tree-level $V_{\bar{K}^*\pi,\bar{K}^*\pi}$, 
or to the box diagram of Fig.~\ref{eleven_boxB} $-$ about $1\%$ for the real part or 
$3\%$ for the imaginary compared to $t_{11b}$ $-$ which gives an idea of the 
relative weight of the pseudoscalar exchange. One should note that we also have 
a diagram in which the vector exchange appears to the left and the pseudoscalar 
exchange appears to the right in Fig.~\ref{eleven_boxA}, which would double its 
contribution but, although smaller than the contribution of Fig.~ \ref{eleven_boxB}, 
we also have an extra contribution of the type of $t_{11b}$ coming from 
Fig.~\ref{eleven_boxB} with $\rho K$ intermediate state, so the corrections 
from pseudoscalar exchange are really small. If we look at $t_{12a}$ and $t_{12b}$ 
the effect seems to be relatively larger: $10\%$ for the real part and 
$44\%$ for the imaginary part. But, if one compares the imaginary part with the 
real part is only $16\%$ correction. Once again, we would double the strength of this 
mechanism by exchanging the $\pi$ and $V$ exchange in Fig.~\ref{twelve_boxA}, 
but we also double the strength of Fig.~\ref{twelve_boxB} by adding $K^*\pi$ in the 
intermediate state. We should also note that the relatively larger corrections 
found in the case of the $\bar{K}^*\pi\to \rho\bar{K}$ transition in Fig.~\ref{Boxtwelve} 
affect an amplitude which, as seen in Appendix \ref{appA}, has a strength $1/4$ 
of the diagonal $\bar{K}^*\pi\to\bar{K}^*\pi$, $\rho\bar{K}\to \rho\bar{K}$ transitions, 
as a consequence of which there is a small mixing of $K ^*\pi$ and $\rho K$ which is 
not affected by the maximum $16\%$ correction to the $K^*\pi\to \rho K$ transition term 
that has a strength of $1/4$ of the diagonal ones. 

The contribution of the box diagram of Fig.~\ref{FigB4} with the exchange of two pions 
given by the $t_{10}$ in Table \ref{tab2} is also very small, and so is that 
of the similar box diagram for $K^*\pi\to K^*\pi$ with $\rho\bar{K}$ intermediate state.

\end{document}